\documentstyle[12pt,cites,epsf,rotate,amssymb,amsfonts]{article}
\begin{document}

\title{\bf Renormalization of the Optical Response of Semiconductors by
Electron-Phonon Interaction }
\author{MANUEL CARDONA\\[18pt]
Max-Planck-Institut f\"ur Festk\"orperforschung, \\
Heisenbergstr.\ 1, 70569 Stuttgart, Germany \\
cardona@kmr.mpi-stuttgart.mpg.de}
\date{August 7, 2001}
\maketitle
\vspace{-0.5cm}
\begin{center}
\end{center}

\noindent In the past five years enormous progress has been made
in the {\em ab initio} calculations of the optical response of
electrons in semiconductors. The calculations include the Coulomb
interaction between the excited electron and the hole left behind,
as well as local field effects. However, they are performed under
the assumption that the atoms occupy fixed equilibrium positions
and do not include effects of the interaction of the lattice
vibrations with the electronic states (electron-phonon
interaction). This interaction shifts and broadens the energies at
which structure in the optical spectra is observed,  the
corresponding shifts being of the order of the accuracy claimed
for the {\em ab initio} calculations. These shifts and broadenings
can be calculated with various degrees of reliability using a
number of semiempirical and {\em ab initio} techniques, but no
full calculations of the optical spectra including electron-phonon
interaction are available to date.

This article discusses experimental and theoretical aspects of the
renormalization of optical response functions by electron-phonon
interaction, including both temperature and isotopic mass effects.
Some of the theoretical techniques used can also be applied to
analyze the renormalization of other response functions, such as
the phonon spectral functions, the lattice parameters, and the
elastic constants.

\section{Introduction}
The temperature dependence of macroscopic properties of crystals
(i.e., specific heat \cite{einstein,nernst,debye}, thermal
expansion \cite{grun}, thermal conductivity \cite{dong,einst},
elastic constants \cite{karch}) played an important role in the
early development of the quantum theory of solids
 \cite{einstein,nernst,debye,grun}.  More recently, the temperature
dependence of microscopic properties, involving spectroscopic
elementary excitations such as those responsible for optical
absorption and reflection \cite{fan,allen9} and phonons \cite{deb}
have been discussed both experimentally and theoretically. The
theoretical treatments have ranged from semiempirical ones
\cite{einstein,nernst,debye,grun,fan,allen9} to state-of-the-art
{\it ab initio} calculations \cite{deb,king}.

During the past five years considerable progress has been made in
calculating {\it ab initio} the linear response functions, i.e.,
the complex dielectric functions of semiconductors and insulators
\cite{rohl,bene,alb}. These calculations, which include the
Coulomb interaction between the excited electron and the hole left
behind in the valence band (the so-called excitonic interaction)
and local-field effects, represent an enormous improvement over
the early semiempirical calculations \cite{brust,cohen}. They
nevertheless implicitly assume that the constituent atoms occupy
fixed positions and thus neglect the effects of the lattice
vibrations on the electronic structure, i.e., the electron-phonon
interaction. Such effects have been calculated for selected
transition energies (at so-called interband critical points, CP)
\cite{fan,allen9,king}, but calculations of the full spectrum of
electronic interband transitions including electron
phonon-interaction effects are yet to be performed.

Considerable  effort has been devoted to experimental studies of
the effect of external perturbations on the dielectric function of
solids. Among these perturbations we mention static strain
\cite{etch17,ronn}, electric fields \cite{last}, magnetic fields
\cite{ali}, and sublattice displacements corresponding to optical
phonons with ${\bf q} = 0$ \cite{ca21}. The latter determine the
frequency dependence of Raman spectra. The contribution of a
crystal surface to the linear optical response has also been
profusely studied. Several articles in these proceedings deal with
this subject, both experimentally and theoretically, the
theoretical work being not as advanced as that mentioned above for
the bulk. In any case it does not include {\it ab initio} the
effects of the electron-phonon interaction either. Nor do the
rather simplified calculations which have been performed for the
effect of strain \cite{etch17,etch22,theo} or electric fields
\cite{hughes} on the dielectric function.

 The effect of ``temperature'' on optical spectra of
crystals results from the interaction of the electronic states
with the vibrations of the lattice (electron-phonon interaction),
an effect which does not vanish even for $T = 0$. According to
quantum mechanics, the vibrational amplitude does not vanish at $T
= 0$: the so-called zero-point vibrational amplitude remains and
is responsible for energy renormalizations up to 100 meV, of the
order of the accuracy claimed for state-of-the-art {\it ab initio}
calculations of critical point energies in such spectra
\cite{rohl,bene,alb,card}.

Zero-point vibrational amplitudes should vanish, in a {\em
gedankenexperiment}, when the ionic masses of a crystal become
infinitely large. In the past 12 years, many crystals with
variable isotopic compositions have become available. It has
therefore become possible to measure the dielectric functions
$\epsilon(\omega)$ of crystals with different isotopic masses, to
extrapolate them to infinite values of $M$ and thus obtain
experimentally the bare values of transition energies
\cite{card,last26,coll,etch}. It may sound paradoxical that one
can obtain \underline{bare}, i.e., unrenormalized, parameters from
experiment since only renormalized parameters, including all
self-energy corrections such as those due to electron-phonon
interaction, are observable. However, as pointed out by Allen
\cite{allen29}, the procedure to obtain unrenormalized parameters
just described is only an approximate one. It requires assumptions
which lead to a simple dependence of the zero-point
renormalization on isotopic mass and temperature, in particular
the truncation of the electron-phonon perturbation theory series
after the second-order terms.

In this paper we describe the three contributions which lead to
the temperature dependence of optical transition energies
(including zero-point effects) in crystals, up to second-order
perturbation theory. They are:

\begin{enumerate}
\item[{\it i})] The effect of the first-order electron-phonon
interaction Hamiltonian to second-order in perturbation theory.
These are the so-called Fan terms \cite{fan}.

\item[{\it ii})] The effect of second-order electron-phonon
interaction in first-order perturbation theory \cite{ant}. These
are Debye-Waller terms, sometimes also referred to as Yu-Brooks
terms \cite{allen9}.

\item[{\it iii})] The effect of thermal expansion, including
zero-point renormalization of lattice parameters, coupled with the
dependence of the transition energies on such parameters
\cite{fan}.
\end{enumerate}

We shall discuss simple algebraic expressions which have been used
to fit the measured dependences of transition energies (also
called gaps or critical points, CP's) on temperature, together
with the related effects of isotopic mass on CP's. The effect
mentioned above under ($i$) is described in terms of complex
self-energies $\Sigma = \Sigma_r + i~\Sigma_i$, which encompass
not only energy shifts ($\Sigma_r$) but also broadenings of the
optical features (Lorentzian-HWHM = $-\Sigma_i$). The effects
$(ii)$ and $(iii)$ correspond to real energy shifts of CP's or
energy gaps. Emphasis is placed in this article on a method of
obtaining zero-point renormalization through extrapolation to zero
temperature of the linear part of the shift of CP energies with
temperature usually found at high temperatures. Extension of these
phonon renormalization to other more macroscopic physical
properties, such as lattice parameters (needed also for the
evaluation of the $(iii)$ terms), elastic constants, and phonon
frequencies, is also discussed.

\section{Renormalization of Electronic States through Electron-Phonon Interaction}

The renormalization of excitation energies by lattice vibrations
(or other perturbing agents such as disorder) is represented, in
the many-body theory of solids, by a frequency (= energy for
$\hbar = 1$, a value of $\hbar$ which will be assumed implicitly
throughout this paper) and wavevector dependent self-energy
$\Sigma(\omega$,{\bf k})= $\Sigma_r(\omega$,{\bf k}) + $i$~
$\Sigma_i(\omega$,{\bf k}). While the \underline{bare} excitation
energy $\omega$({\bf k}) is real, i.e., the excited ``particles''
with a well-defined real wavevector {\bf k}
 have an infinite lifetime, the imaginary part of the self-energy
adds an imaginary component to the excitation energy and the
excited ``particle'' acquires a finite lifetime $\tau$ for a real
value of {\bf k}, i.e., it becomes a ``quasiparticle''. This
lifetime of the corresponding electron-hole  excitation is given
by the uncertainty-principle-like expression:

\begin{equation} 
\tau = \left(- 2 ~\Sigma_i\right)^{-1}~ ,
\end{equation}

\noindent the minus sign resulting from the convention chosen for
the sign of $\Sigma_i~ (\Sigma_i\leq 0)$. We are  concerned here
with frequency renormalizations induced by phonons of frequency
$\Omega$({\bf q}) and harmonic vibrational amplitude
$u(\Omega$,{\bf q}). Because the time average of $u$ is zero (in
the harmonic approximation, perturbation terms of first-order in
$u(\Omega$,{\bf q}) vanish in $\Sigma(\omega$,{\bf k})). Such
terms are responsible for a broad, incoherent (i.e., that does not
correspond to a single value of {\bf k}) background which will not
concern us here, except in so far as it has been incorrectly
interpreted sometimes to generate a finite value of
$\Sigma_i(\omega$,{\bf k}) \cite{cicca}.

\subsection{Thermal expansion effects}

As already mentioned, the calculation of $\Sigma(\omega$,{\bf k})
is broken up into two parts. One of them involves a perturbation
calculation keeping the solid at constant volume (i.e., keeping
the shape of the primitive cell and the Brillouin zone, BZ,
constant). However, the experiments are not performed at constant
volume, but at constant pressure. Hence, we must add to the
perturbation results just mentioned the effect of thermal
expansion at constant pressure (usually atmospheric, for practical
purposes zero):

\begin{equation} 
\Delta\omega_0 = \left( \frac{\Delta V(T)}{V_0}\right)_p \cdot~
a~= \left(\frac{\Delta V(T)}{V(0)}\right)_p \cdot~B_0 \cdot~
\left(\frac{\partial\omega_0}{\partial p}\right)_T
\end{equation}

\noindent where $\Delta V(T) = V(T)-V_0$ is the change in volume
induced by the temperature $T$ and $a$ is the change in transition
energy $\omega_0$ induced by the change in volume, the so-called
hydrostatic deformation potential of the CP under consideration.
Usually one measures the dependence of $\omega_0$ on pressure, in
which case $\Delta\omega_0$ must be obtained from the r.h.s. of
(1) using the bulk modulus $B_0$ and the pressure dependence of
$\omega_0$ represented by $(\partial\omega_0/\partial p)$.

The thermal expansion $(\Delta V(T)/V_0)_p$ results from {\em
anharmonic} terms in the expansion of the crystal energy vs.
atomic displacement, whereby only odd terms in this expansion
contribute (or combinations of odd and even terms, with an odd
number of odd terms): The even terms average to zero and do not
change the equilibrium position of the atoms. We shall keep for
our discussion only the lowest (third) order anharmonic terms
which lead to a linear dependence of $\Delta V(T)$ on $T$ at high
temperature, i.e., to a $T$-independent thermal expansion
coefficient in this $T$ range. These terms are conveniently
represented by the {\em mode Gr\"uneisen parameter}
$\gamma_n$({\bf q}) defined as:

\begin{equation} 
\gamma_n({\bf q})~=~ - \frac{V_0}{\Omega_n({\bf q})}~\left(
\frac{\partial\Omega_n({\bf q})}{\partial V}\right)_{V_0}
\end{equation}

\noindent where {\bf q} is the wavevector of a generic phonon and
$n$ its branch index. Using themodynamic relations, and minimizing
the free energy at ``zero pressure'' and temperature $T$, one
finds \cite{pavone,deber}:

\begin{equation} 
 V(T) = V_0~+~ \frac{1}{2B_0} ~ \sum_{n,{\bf q}}
~\Omega_n({\bf q})~ \gamma_n ({\bf q})(2n_B+1)
\end{equation}

\noindent where $n_B$ is the Bose-Einstein statistical factor:

\begin{equation} 
 n_B(\Omega,T)~=~ \left[e^{\xi}-1\right]^{-1}~; \xi =
\frac{\Omega}{T}~ ,
\end{equation}

\noindent with $T$ in the same units as $\Omega$. Note that
Eq.~(4) allows us to calculate not only the temperature dependence
of the crystal volume, but also the zero temperature
renormalization, by setting $n_B=0$
 \cite{pavone,deber,london,busch}. In elemental crystals this
zero-point renormalization is proportional to $M^{-\frac{1}{2}}$,
where $M$ is the average isotopic mass.

In order to evaluate the thermal expansion contribution to
$\Delta\omega_o$ (including zero-point effects) with Eq.~(2) one
can use either experimental values of $(\Delta V(T)/V_o)$
 \cite{sozont,reeber} or those calculated with Eq.~(4). The former are
obtained rather accurately by means of X-ray diffraction. For the
latter procedure one can use either experimental values of
$\gamma_n$({\bf q}) \cite{wein} or values calculated using {\em ab
initio} band structure techniques \cite{pavone,deber}. One should
mention, at this point, that the $\gamma_n$({\bf q}), which
according to (4) should have an average value of about +1, is
negative $(\simeq-1)$ for the TA phonons at the edge of the BZ of
most germanium and zincblende-type semiconductors (exceptions:
diamond \cite{wein} and, curiously enough, CuI
\cite{wein,serrano}). When these negative values are entered into
Eq.~(2) they lead to anomalous negative expansion coefficients at
temperatures close to the frequencies of the corresponding
phonons. At much lower temperatures, the expansion coefficients
become usually (but not always \cite{deber}) positive again. This
results in a very flat dependence of $V(T)$ vs. $T$ for $T$ below
the TA-phonon frequencies \cite{reeber,lyon}. The oscillations due
to the sign reversals of $\gamma_n$({\bf q}) are easy to see in
the expansion coefficient vs. $T$ which corresponds to the {\it
derivative} of $V(T)$ vs. $T$. They are hardly noticeable in
$\Delta V(T)$ except that the behavior of $\Delta V(T)$ vs. $T$ is
very flat at low temperatures. It is therefore very difficult to
see the corresponding effects in the contribution to
$\Delta\omega_0$ given in Eq.~(2), a fact which, however, makes
easier the analysis of data on the temperature dependence of
electronic CP energies.

\subsection{Electron-phonon interaction}

The direct electron-phonon interaction terms, defined for a
temperature independent volume, are responsible for the ``explicit
effect'' of temperature on the electron energies. The term
``implicit effects'' is applied to the thermal expansion
contribution discussed in 2.1. The two terms displayed as Feynman
diagrams in Fig.~1 contribute to the phonon renormalization of
one-electron states to the same order in perturbation theory as
that used in 2.1.
 Electronic excitations involve an excited electron in
the conduction band and a hole left behind in the valence band. It
is easy to show that the renormalization can be obtained as the
difference of two sets of diagrams of the type in Fig.~1, the
electron diagram minus the hole diagram, provided one neglects
vertex terms which connect electron and hole lines \cite{allen41}.
Diagram $(i)$, representing the first-order perturbation
Hamiltonian taken to second order in perturbation theory, leads to
a complex self-energy whose imaginary part  $\Sigma_i$ represents
the Lorentzian width (HWHM) of the renormalized state. This
$\Sigma_i$ only vanishes at the uppermost valence state and at the
lowest conduction state, i.e., for the lowest absorption edge
(which can be either direct, like in GaAs, CdS, CuCl and GaN, or
indirect, like in Ge, Si, diamond and GaP).
 The absorption spectra
corresponding to this edge can, therefore, be very sharp at low
$T$ \cite{karai}. Diagram $(ii)$ represents the second-order
electron-phonon perturbation Hamiltonian, i.e., the second
derivatives of the Hamiltonian with respect to the vibrational
displacements, taken only to first order in perturbation theory
and thus leading to simple frequency shifts without imaginary
energy components. It is not uncommon to find in the older
literature frequency renormalizations corresponding to only one of
the diagrams in Fig.~1. Since 1983, however, both components of
the electron-phonon interaction have been usually evaluated and
added \cite{allen9,zoll}.

\begin{figure}[htb]
\epsfxsize=.45\textwidth \centerline{\epsffile{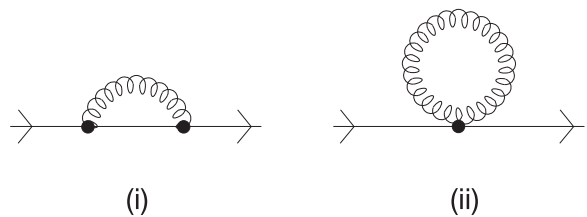}}
\caption[]{Electron-phonon interaction terms that contribute to
the renormalization of electronic states. $(i)$ Fan terms
corresponding to a complex self-energy. $(ii)$ Debye-Waller terms
leading to a real energy correction.} \label{pssa1}
\end{figure}

For elemental crystals the electronic energy shift induced by the
term $(ii)$ can be written as \cite{allen9}:

\begin{equation} 
\Delta\omega(T)~=~ \sum_{n,{\bf q}} ~ \frac{A_{DW}(n,{\bf
q})}{\Omega(n,{\bf q} )M} ~ (2n_B+1)
\end{equation}
\vspace{0.5cm}

\noindent where $A_{DW}(n,${\bf q}) is a real coupling constant
specific to the electronic state under consideration and $M$ the
mass of the element at hand. In a {\bf k}-dependent
pseudopotential representation, Eq.~(6) can be evaluated by
multiplying the pseudopotential form factors by Debye-Waller
factors \cite{ant}. This procedure reduces simply to performing
band structure calculations with two sets of pseudopotentials, the
bare one and that screened by temperature-dependent Debye-Waller
factors. This is the reason why calculations are often performed,
using {\it Occam's razor},
 only for the $(ii)$ terms, neglecting $(i)$
\cite{yu}, a procedure known nowadays to be unjustifiable.
Actually, when using a pseudopotential or any kind of truncated
Hamiltonian for the calculation, the separate real parts of $(i)$
and $(ii)$ do not have an independent meaning as they depend on
the type of truncation at hand. Only their sum is physically
meaningful.

The self-energy terms $(i)$ of Fig.~1 can also be represented by
an equation of the type of Eq.~(6) with a complex coupling
constant $A_{SE}(n$,{\bf q}) which corresponds to the typical
second-order perturbation expressions, with squared matrix
elements in the numerator and an energy denominator with an
infinitesimal imaginary convergence term in it \cite{allen9}. Its
numerical evaluation is considerably more complicated than that of
the Debye-Waller terms $(ii)$, since it involves not only a sum
over all phonon states, but also one over intermediate electronic
states. The early analytical calculation by Fan \cite{fan}, while
pioneering, is therefore highly unreliable.\footnote{{\em Fan
obtained reasonable agreement with experimental data available to
him, in spite of not having considered the $(ii)$ terms. This is
not unusual in the case of theories constructed so as to explain
extant experimental results.}} Fan suggested a procedure to relate
the $(i)$ to the $(iii)$ terms by assuming parabolic band edges in
order to perform analytically the integration of Eq.~(6). Two
parameters, corresponding to the deformation potentials of the
conduction and valence band edges, are then introduced. Their
difference determines the $(iii)$ terms whereas the properly
weighted difference of their squares determines the $(i)$ terms.
The $(ii)$ terms are ignored. This procedure, if correct, allows
the separate determination of the hydrostatic deformation
potentials of the valence and conduction bands as done in
Ref.~[43]. However, as already mentioned, the results of this
procedure are not trustworthy.

Note the similarity between Eqs.~(6) and (4). At temperatures
higher than the average phonon frequency (sometimes called the
Debye frequency), they become linear in $T$. This fact, which
follows from our assumption of terms only up to second order in
the corresponding perturbation expansion \cite{fan,allen9,zoll}
 is often corroborated by
experiments up to temperatures somewhat below the melting point.
For $T\rightarrow 0,$ Eqs.~(4) and (6) tend to finite
contributions to the zero-point renormalization. The
high-temperature expansion of Eq.~(6) can be written as

\begin{equation} 
\Delta\omega = \frac{1}{M}~\left[\Big\langle \frac{A}{\Omega}
\Big\rangle +2\Big\langle \frac{A}{\Omega^2}\Big\rangle T\right]~,
\end{equation}

\noindent where the brackets represent average values. A similar
expression applies to the high-temperature limit of Eq.~(4).
Considering the fact that in elemental crystals all frequencies
$\Omega$ are proportional to $M^{-\frac{1}{2}}$, we conclude from
Eq.~(7) that in the high-$T$ limit $\Delta\omega$ is independent
of $M$ (classical limit). For $T\rightarrow 0$ (quantum limit)

\begin{equation} 
\Delta \omega \propto M^{-\frac{1}{2}} ~ .
\end{equation}

Equations (7,8) apply to all three components $(i,ii,iii)$ of
$\Delta\omega$. If we are able to observe a linear region in
$\Delta\omega(T)$ at large $T$, the extrapolation of this
asymptotic linear behavior to $T=0$ enables us to estimate the
zero-point renormalization of the corresponding critical point
(CP) energy $\Delta\omega$ ($T=0$).

Figure 2 displays the temperature dependence of the energy of the
lowest (indirect) gap of germanium and the linear high-$T$
asymptote extrapolated to $T=0$ \cite{laut,thur}. This
extrapolation leads to the estimate $\Delta\omega$ $(T=0)$ = -53
meV for the zero-temperature phonon renormalization of the gap
energy at 0 K. Using Eq.~(8) we can estimate a difference of 2.2
meV between the corresponding gaps of $^{76}$Ge and $^{70}$Ge,
which compares well with the measured value of 2.2 meV
\cite{parks}. Figure 2 displays not only the experimental points
for $E_g(T)$ but also the calculated sum of the $(i)$ and $(ii)$
contributions (dashed line) and  this sum plus the thermal
expansion contribution $(iii)$ (solid line) \cite{laut}.

\begin{figure}[hbt]
\epsfxsize=.35\textwidth
\centerline{\epsffile{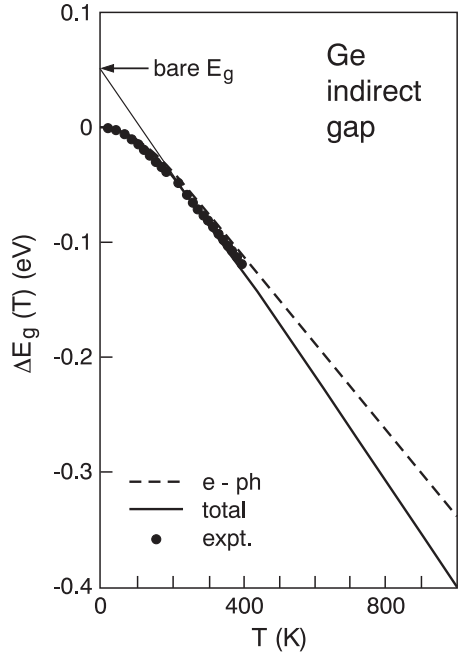}}
\caption[]{
 Temperature dependence of the energy of the lowest
indirect gap of germanium. The dashed line represents the effect
of electron-phonon interaction. The thick line represents this
effect plus the thermal expansion effect. The thin line represents
the asymptotic extrapolation to the bare gap at $T = 0$. The
points are experimental \cite{laut}.} \label{pssa2}
\end{figure}

We should  recall that the direct gap $E_0$ of Ge is not its
lowest gap, there being the indirect $E_g$ gap ($\Gamma-L$) just
discussed about 150 meV below. Hence, the zero-point self-energy
of the direct gap $E_0$ should have an imaginary part which should
be small because this gap is not far from being the lowest. From
Fig.~1 in Ref.~48 we estimate $-\Sigma_i\simeq$ 0.5~meV at
$T\simeq 0$ K, much smaller than $\Sigma_r$ = 70 meV \cite{parks}.

We have, thus far, lumped all contributions $(i),(ii),(iii)$ to
$\Sigma$ into a single one. In order to illustrate that the three
contributions have the same asymptotic behaviors, we display in
Fig.~3 the measured change in lattice parameter $a_0(T)$ of
diamond with temperature (related to $\Delta V(T)$ in a
straightforward way) \cite{temp}. The asymptotic linear
extrapolation leads to a zero-point renormalization $\Delta
a_0/a_0$ = 0.374 which, using Eq.~(8), allows us to predict a
difference between the lattice constants of $^{12}C$ (larger) and
$^{13}C$ equal to $1.5\times 10^{-4}$, in excellent agreement with
the experimental results and {\it ab initio} calculations
\cite{hollow,pavone}. Notice that the one-oscillator fit of Fig.~3
leads to an average phonon temperature $\Omega=$ 1460K, which
compares well with the corresponding value for the specific heat
(1450K \cite{debye}). The three oscillator fits of \cite{reeber}
leads to a dominant frequency $\Omega =$ 2137K, rather close to
that of a two-oscillator fit of the specific heat (1884 K
\cite{nernst,debye}).

\begin{figure}[htb]
\epsfxsize=.35\textwidth
\centerline{\epsffile{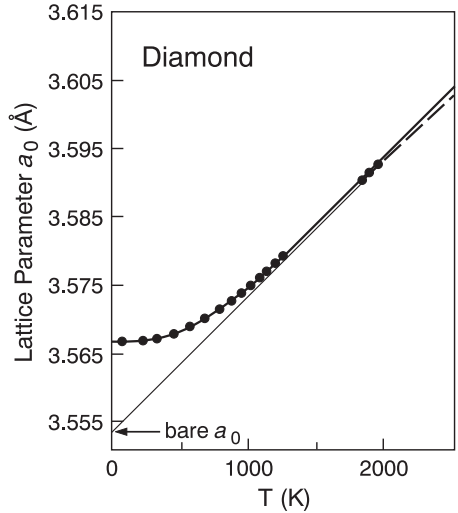}}
\caption[]{Lattice parameter $a_0$ of natural diamond vs.
temperature. The points represent schematically the experimental
data (up to 2000K). The thin line is the high temperature
asymptote which extrapolates to the bare $a_0$ for $T\rightarrow
0$ (horizontal arrow). The thick curve represents a
three-oscillator fit whereas the dashed curve (only plotted for $T
>$ 2000K) corresponds to a one-oscillator fit. See \cite{reeber} and
text.} \label{pssa3}
\end{figure}

Concerning the dependence of physical parameters on isotopic mass,
in this section we have confined ourselves to elemental crystals:
only the mass of one element appears in Eqs.~(7,8). In the case of
binary or more complex crystals, that dependence is different for
each one of the constituent masses and Eq.~(7) becomes
considerably more complicated. The experimental \cite{gob} and
theoretical \cite{garro} information concerning dependences on
isotopic masses for compound semiconductors is rather scarce. It
can, nevertheless, be very interesting. In the case of the copper
halide CuCl, for instance, the lowest gap ($E_0)$ increases with
increasing Cl-mass (a trend which agrees with Fig.~2) whereas it
anomalously decreases with increasing Cu-mass \cite{gob,garro}.

\section{Approximate Expressions for the Dependence of CP Energies
on Temperature}

A number of empirical \cite{varsh} and semiempirical
 \cite{vina,mano,pass} analytical expressions have been proposed to represent
the measured $\Delta\omega_0(T)$ and the corresponding imaginary
part of the self-energy. The most popular fully empirical
expression, proposed by Varshni, \cite{varsh} is:
\begin{equation} 
\Delta \omega_0 (T) = ~ \frac{\alpha T^2}{\beta + T}~ .
\end{equation}
At high $T$, Eq.~(9) becomes linear in $T$, in agreement with the
discussion of Sect.2. For $T\rightarrow 0$, however, it becomes
quadratic in $T$, leading to a stronger variation with $T$ than
observed experimentally (see Fig.~2 in Ref.~56 for GaAs).

The semiempirical expressions usually capitalize on the
statistical factors $(2n_B+1)$ which appear in Eqs.~(4) and (6),
replacing them by average values. In Ref.~54 one average value,
corresponding to a single Einstein oscillator as used in the
theory of the specific heat \cite{einstein}, is employed. In
Refs.~43 and 55 two oscillators are used, paralleling the improved
theory of the specific heat of Nernst and Lindemann \cite{nernst}.
Like in the case of the specific heats \cite{debye}, electronic
self-energies due to electron-phonon interaction at low $T$
deviate from the single oscillator behavior. Two oscillators
\cite{nernst,mano} improve matters a bit but, $n_B$ implying an
exponential dependence on $T$, cannot reproduce the $T^3$ behavior
found experimentally for the specific heat of insulators. It is
easy to prove that Eqs.~(4), as well as (6), also lead to a low
temperature behavior proportional to $T^3$. This is accomplished
in Eq.~(4) by replacing the sum by an integral over energy, which
requires the introduction of the density of phonon states,
proportional to $\Omega^2$ at low $T$. Switching to the
dimensionless variable of intergration $\xi = \Omega/T$ leads to
the factor $T^3$ that we are looking for.\footnote{{\em Note that
if $\gamma_n$({\bf q}) undergoes the changes in sign mentioned
above, the range of $T^3$ behavior for the thermal expansion at
low temperatures will be rather small. However, as already
mentioned, these changes in sign make the behavior of V(T) vs. $T$
very flat at low $T$, thus improving the overall quality of single
oscillator fits, although they cannot describe the resulting
non-monotonic behavior of the expansion coefficient for
$T\rightarrow 0$.}} In order to obtain the same factor with
Eq.~(6), we need some additional reasoning. We must consider the
fact that translational invariance requires that the
electron-phonon coupling constant vanishes for $\Omega\rightarrow
0$: long wavelength acoustic phonons are equivalent to a uniform
translation which does not couple to the electronic states. Hence,
a factor of $\Omega^2$ appears in $A_{DW}$ and $A_{SE}$. The same
factor of $T^3$ is then obtained in Eq.~(4) for $T\rightarrow0$.
We thus see that Eq.~(9), leading to $\Delta\omega_0(T)\propto
T^2$ for $T\rightarrow 0$, cannot be correct either. Several
attempts to remedy this problem, using empirical expressions,
appear in the works of P\"assler \cite{pass}. We illustrate them
in Fig.~4 for the indirect gap of Si. The experimental points fall
above the single oscillator fit of Ref.~54, as expected. Also as
expected, the fit with Eq.~(9) falls well above the single
oscillator fit of Ref.~54. An excellent fit is obtained with the
four-parameter expression:

\begin{figure}[htb]
\epsfxsize=.35\textwidth
\centerline {\epsffile{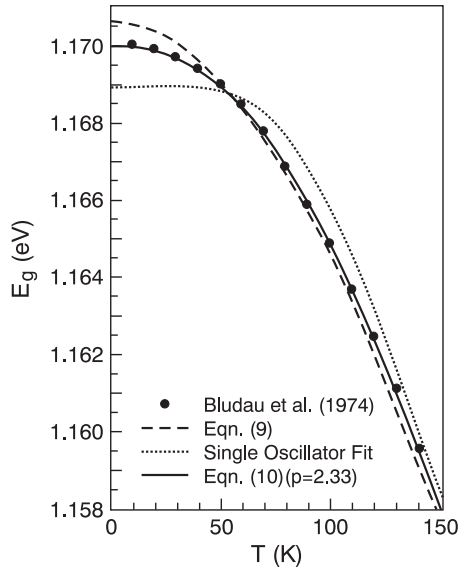}}
\caption[]{Temperature dependence of the indirect gap of silicon
$E_g(T)$. The points are experimental, the solid line
 represents a fit with Eq.~10. Other details are given in the
 figure and in the text. Adapted from \cite{pass}.}
\label{pssa4}
\end{figure}

\begin{equation} 
E_g(T) = E_g(0)~ - \frac{\alpha\Theta}{2}~
\left[^p\sqrt{1+\left(\frac{2T}{\Theta}\right)^p}-1\right]
\end{equation}

\noindent for $p = 2.33$. Equation (10) leads to a linear
$T$-dependence for $T\gg\Theta$, as it should. Obtaining a good
fit with an empirical or semiempirical expression is very helpful
for the precise determination of the linear asymptotic behavior
which, as discussed in Sect.~2, can be used to find the zero-point
gap renormalization.

An excellent fit to $E_g(T)$, such as that provided by Eq.~(10)
(with $p$ = 2.33, $\Theta = 405.6$ K, $\alpha = 0.3176$~ meV
$\cdot$ K$^{-1}$), can be used to circumvent the problem of
determining the high temperature asymptote, and thus the
zero-point renormalization mentioned above. Using Eq.~(10) we
obtain for the $T=0$ renormalization:

\begin{equation} 
\Delta E_0 (T = 0) ~=~ -\frac{\alpha\Theta}{2} ~ .
\end{equation}

\noindent Replacing into Eq.~(11) the value $\alpha = 0.3176$ $\rm
meV \cdot K^{-1}$, and $\Theta$ = 405.6 K given in [48], we find
$\Delta E_0 (T=0) = -64$ meV for the indirect gap of Si. From this
value of the zero-point renormalization  one can estimate with
Eq.~(8) the isotope shift of the indirect gap of silicon, between
natural silicon ($M = 28.0856$) and isotopically pure $^{28}$Si,
to be 0.12 meV,  in good agreement with the experimental results
(0.094 meV) \cite{karai}.

We show in Table I the values of the zero-point renormalization of
the lowest gaps (plus the lowest {\em direct} gap of Ge) of
several diamond and zincblende-type semiconductors, together with
the corresponding values of $dE/dM$. The first column under
$\Delta E(T=0)$ and $dE/dM$ represents experimental data obtained
by the linear extrapolation technique and by isotopic
substitution, respectively (an exception is the case of diamond,
for which the calculated value  of $\Delta E(T=0)$ \cite{zoll} is
given). The available experimental data for diamond do not allow a
reliable extrapolation to $T=0$, the experimental value of
$dE_g/dM$ listed in Table I is from \cite{collins}. The numbers in
the second column under $\Delta E(T=0)$ (and $dE/dM$) were
estimated from the ``experimental'' values of $dE/dM$ (and $\Delta
E(T=0))$ using the $M^{-\frac{1}{2}}$ dependence, as discussed
above. The values of $\Delta E(T=0$) are rather similar for most
materials in the table, with the exception of those for diamond,
which are very large, and the rather small ones listed for the
copper halides (to be discussed in Sect.~4.2). We note that
several electron-phonon coupling constants of diamond are known to
be considerably larger than those for other zincblende-type
semiconductors \cite{ca58}, a fact which we conjecture it arises
from the absence of p-like electrons in the core of the carbon
atoms. Correspondingly, the measured $dE_g/dM$ is also rather
large for diamond \cite{collins}.

\section{Lowest Direct Gap $\rm E_0$ of compound Semiconductors:
Temperature and Isotope Effects}

Information on the temperature dependence of the lowest gaps of
many semiconductors can be found in standard compilations of
semiconductor properties (e.g., the Landoldt-B\"ornstein Tables
and Ref. [47]). We shall discuss in this section a few cases for
which isotopic data are also available.

\subsection{GaAs and ZnSe}
The temperature dependence of  $E_0$ for GaAs is rather similar to
that for germanium, as expected from the similarity in their band
structures and lattice dynamics. The parameters for the best fit
with Eq.~(10) are $\Theta$ = 226 K, $\alpha = 0.473$ meV
$\cdot$K$^{-1}$, $p$ = 0.251 \cite{passl}. Using these parameters
one predicts with Eq.~(11) $\Delta E_0 (T = 0) = -53$ meV, a value
very close to that given in Table I  for Ge. Unfortunately,
arsenic has only one stable isotope ($^{73}$As) so that only the
effect related to the two isotopes of Ga, $^{69}$Ga and $^{71}$Ga,
can be experimentally investigated. Common sense, and also
pseudopotential calculations \cite{garro}, suggest that the effect
of varying the Ga mass should be nearly the same as that of
varying the As mass (per unit atomic mass). The effect of the
$^{69}$Ga-$^{71}$Ga substitution, realized experimentally, can
thus be predicted by dividing the prediction of  Eq.~(8) for
germanium by 2. We thus predict for the derivative of $E_0$ with
respect to the Ga mass:

\begin{equation} 
\frac{\partial E_0}{\partial M_{Ga}} = \frac{1}{4\times70}
 \times 53 = 0.19~ meV/amu
\end{equation}

This value is a factor of two smaller than the measured one
(0.39~meV/amu) and also than the value obtained from the
pseudopotential calculations (0.43~meV/amu) \cite{garro}. The
reason for this discrepancy is not known but it may lie on having
taken too small a temperature range for the fit in
Ref.~\cite{passl}. Indeed, if we use the data for $E_0(T)$ given
in Fig.~2 of Ref. \cite{thur} for GaAs we find $\Delta E_0(T=0)=
$-90 meV, a value which leads to $dE/dM_{Ga}$, in rather good
agreement with the experimental result (see Table I).

For ZnSe, the linear extrapolation of $E_0(T)$ gives a zero-point
renormalization $\Delta E_0(T=0) = -65$ meV \cite{garro}.
Pseudopotential calculations \cite{garro} yield a stronger
electron-phonon contribution to the isotope effect for Se
$(\partial E_0/\partial M_{Se}=0.26)$ than for Zn ($\partial
E_0/\partial M_{Zn}=0.19$),\footnote{{\em Note that this trend
continues for CuBr where $\partial E_0/\partial M_{Cu}$ even
becomes negative \cite{gob}}} although the difference is
compensated, at least partially, by the corresponding thermal
expansion effects \cite{deber,garro}. The measured values of
$\partial E_0/\partial M_{Zn}$ = 0.214 and $\partial E_0/\partial
M_{Se}$ = 0.216 are indeed nearly equal (see \cite{gob,gobel} and
Table I). The value estimated with Eq.~(8) assuming that both
derivatives are equal is 0.24 meV, in reasonable agreement with
the experimental results.

Notice that the thermal expansion contribution $\Delta
E_{th}(T=0)$ shown in Table I for GaAs is about 40\% of the total
gap renormalization $\Delta E_0(T=0)$ and has the same sign. In
the case of GaP, however, the thermal expansion contribution to
the indirect gap is only 4\% of the total effect and has the
opposite sign. This apparent anomaly is related to the large value
and the negative sign of the deformation potential of direct $E_o$
gaps and the small value plus sign reversal which takes place for
the $\Gamma\rightarrow\Delta$ indirect gaps $E_g$. Similar effects
can be seen in Table I for $\Delta E_0(T=0)$ of Ge as compared
with $\Delta E_g(T=0)$ of Si. For the indirect gap of germanium
$E_g$, the sign of the thermal expansion renormalization is the
same as that of $E_0$ and its absolute value is somewhat smaller
(25\%) as corresponds to a $\Gamma\rightarrow L$ indirect gap.

\subsection{The cuprous halides}
CuCl, CuBr and CuI (I-VII compounds) crystallize, under normal
conditions, in the zincblende structure. However, when compared
with their group IV, III-V and II-VI counterparts, they display a
number of anomalous properties \cite{garro,gobel,ca} which are
usually attributed to a strong hybridization of the halogen $p$
valence bands with the 3d electrons of copper. Among these
properties we mention, as relevant to the present paper, the
increase in $E_0$ with increasing $T$ observed for CuCl and
CuBr.\footnote{{\em For CuI $E_0$ decreases only very slightly
between 0 K and room temperature \cite{ca,schweiz}.}}

\begin{figure}[hbt]
\epsfxsize=.35\textwidth
\begin{minipage}{.6\textwidth}
\centerline{\epsffile{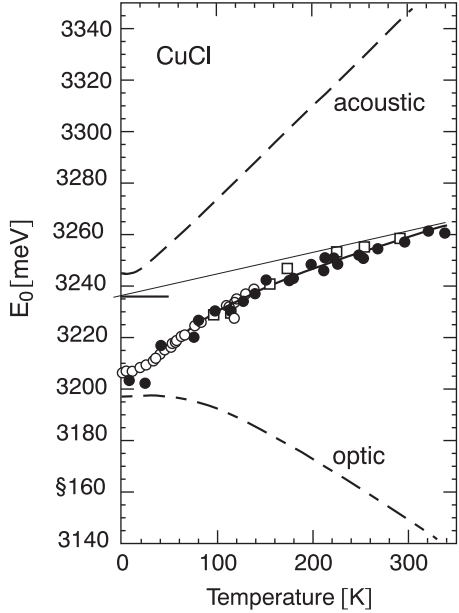}}
\end{minipage}\hfill
\begin{minipage}{.35\textwidth}
\caption[]{Temperature dependence of the $Z_3$ exciton energy
(equivalent to that of the $E_0$ direct gap) of CuCl. The points
are experimental, the thick solid curve represents a
two-oscillator fit (acoustic and optic) while the thin line
indicates the linear asymptote. From \cite{gob}.} \label{pssa5}
\label{pssa5}
\end{minipage}
\end{figure}

Figure 5 displays the anomalous temperature dependence of the
$E_0$ gap of CuCl \cite{gob}. The experimental points show an
increase in gap energy with increasing temperature, with an
increasing slope at low $T$ that begins to decrease at $\sim$100~K
and seems to saturate at $\sim$300~K. This behavior can be
phenomenologically interpreted by means of a two-oscillator fit
\cite{nernst}, provided one assumes that the two oscillators give
electron-phonon contributions of opposite signs: The lower
frequency oscillator, most likely related to Cu vibrations,
contributing a positive $\Delta E_0(T=0)$ and the high frequency
one, related to Cl, giving a negative contribution. These opposite
signs agree with those determined experimentally for the
derivatives of $E_0$ with respect to the copper and chlorine
masses $(\partial E_0/\partial M_{Cu}= -0.076$~ meV/amu, $\partial
E_0/\partial M_{Cl}= +0.36$~ meV/amu, see Fig.~6.) \cite{gob}.

\begin{figure}[h]
\epsfxsize=.30\textwidth
\begin{minipage}[b]{.6\textwidth}
\centerline{\epsffile{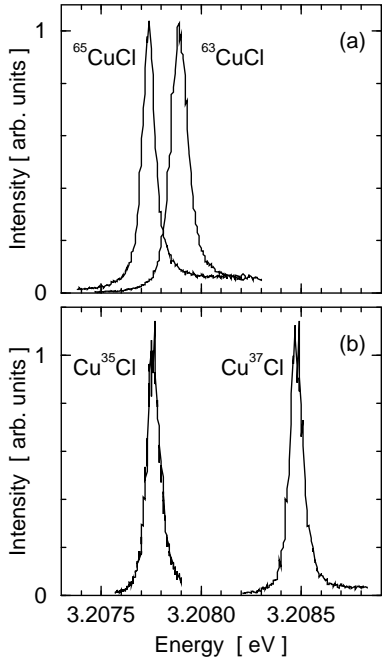}}
\end{minipage}\hfill
\begin{minipage}[b]{.35\textwidth}
\caption[]{Two-photon absorption in the region of the
 longitudinal exciton (LE1) in isotopically modified CuCl measured
 at 2K. The isotopic shifts shown in this figure are expected to
 correspond to those of the direct gap $E_0 + \Delta_0$, which
 should be nearly the same as those of $E_0$. Notice that the
 shifts with increasing mass are negative (i.e., anomalous) for
 Cu and positive for Cl \cite{gob}.}
\label{pssa6}
\end{minipage}
\end{figure}

The question arises as to whether the anomaly just mentioned is
due to the electron-phonon interaction or to the thermal expansion
term. Fortunately, the deformation potential $\partial
E_0/\partial ln V$, and also the thermal expansion, are very small
and the corresponding effect can be neglected for CuCl and
approximately neglected for CuBr, see Table I.  The anomalous
contribution of Cu, and that of the halogen, must then be due
mainly to electron-phonon interaction.

The two-oscillator contributions to $\Delta E_0(T)$ are labeled
``acoustic'' and ``optic'' in Fig.~5. The anomalous acoustic
contribution has been attributed in Ref.~51 predominantly to the
Cu vibrations, the optic contribution, with the usual sign, to
chlorine. An excellent fit to the experimental data is obtained
when adding both contributions. The horizontal line represents the
bare gap as obtained from the two-oscillator fit. A thin straight
line represents the asymptotic behavior at high $T$ and
extrapolates to the bare gap for $T=0$.

A similar anomalous behavior of $E_0(T)$ and $\partial E_0/M_{Cu}$
is found for CuBr \cite{gob,ca,schweiz}. A negative $\partial
E_0/\partial M_{Cu}$ is also found for CuI.\footnote{{\em
Unfortunately, iodine has only one stable isotope} ($^{127}$I).}
However, at least below 300 K for CuI, $E_0(T)$ decreases with
increasing $T$ \cite{ca}. This case is in the process of being
analyzed \cite{schweiz}.

\section{Higher Critical Points: Dependence on Temperature and
Isotopic Mass} We have labeled as $E_0$ the lowest direct critical
point (i.e., energy gap) of a germanium and zincblende-type
semiconductor. In all cases discussed, $E_0$ is found around the
$\Gamma$-point (center) of the Brillouin zone. If it is the
absolute minimum of all transition energies, including indirect
ones (e.g., for GaAs, see Fig.~7, \cite{lauten}), the
corresponding $\Sigma_i$ vanishes.\footnote{{\em Note, however,
that a broadening of the indirect absorption edge of natural
(i.e., isotopically mixed) Si has been recently discovered. It
disappears for isotopically pure $^{28}$Si and thus can be
attributed to isotopic disorder \cite{karai}}.} If not, a finite
value of $\Sigma_i$ is expected (e.g., Ge, $-\Sigma_i=0.5$ meV at
$T=0$, see Sect.~3). The $E_0$ CP at $\Gamma$ is split by
spin-orbit interaction, the second component being known as $\rm
E_0+\Delta_0$. A finite value of $\Sigma_i$ is expected for this
second component since it is not the lowest: it is induced by the
electron-phonon interaction between the coresponding valence state
at the $\Gamma$ point and the top valence band of $\Gamma$, at
nearly the same energy \cite{lauten,yu68,law,aspnes} (($i$) terms
of Fig.~1). In the case of GaAs, $-\Sigma_i$ for $E_0 + \Delta_0$
has been found to be 6 meV at 2K \cite{law,aspnes}. The behavior
of -$\Sigma_i$  with increasing temperature can be estimated by
multiplying 6 meV by $2\langle n_B\rangle+1$, $\langle n_B\rangle$
being the Bose-Einstein factor for the appropriate average phonon
frequency.
\begin{figure}[h]
\epsfxsize=.55\textwidth \centerline{\epsffile{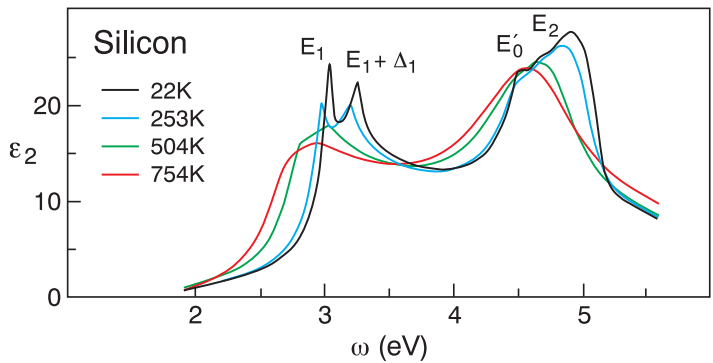}}
\caption[]{Spectra of the dielectric function $\epsilon_2(\omega)$
of GaAs ellipsometrically measured at several temperatures. Notice
the critical points labeled $E_1, E_1+\Delta_1, E'_0,$ and $E_2$.
From \cite{lauten}.}
\label{pssa7}
\end{figure}

$E_0$ and $E_0 +\Delta_0$ correspond to optical absorption
coefficients of the order of $10^4$ cm$^{-1}~(\epsilon_2 \simeq
1)$. With increasing photon energies the absorption coefficient
rises to a maximum near $10^6$ cm$^{-1}$: The corresponding CP is
labeled $E_2$. The $\epsilon_2(\omega)$ spectrum of GaAs in the
region of strong electronic absorption (1.5 to 5.5 eV) is shown in
Fig.~7 for several temperatures. The strongest feature corresponds
to the $E_2$ CP's: it red-shifts and broadens strongly with
increasing $T$. Another feature, labeled $E'_0$, corresponds to
transitions between the $\Gamma_{25'}$ valence band and the
$\Gamma_{15}$ conduction band states \cite{lauten}.

We display in Fig.~8 $\epsilon_2(\omega)$ spectra measured for
GaAs at 20 and 300 K (from Fig.~7) and compare them with recent
{\em ab initio} calculations of Rohlfing and Louie \cite{rohl},
one of the calculated spectra (dashed line) without electron-hole
(exciton) interaction. The measured spectra display clearly the
spin-orbit splitting of the $E_1$ transitions
$(E_1-E_1+\Delta_1)$.  The spin-orbit interaction was not included
in the calculations in order to reduce (by a factor of
$2\times2=4$) the size of the Hamiltonian matrix. The calculations
include, however, an artificial broadening $\Gamma\backsimeq$
0.10~eV so as to smooth out the limited {\bf k}-space integration
mesh. This is the reason why the calculated $E_1-E_1+\Delta_1$
structure, in principle not including lattice vibrations, is
broader than that measured at 20K. Unfortunately, because of the
single isotope nature of arsenic ($^{73}$As) and the rather small
mass difference between the two isotopes of Ga ($^{69}$Ga,
$^{71}$Ga), the experimental accuracy does not
allow for an
extrapolation of the $\epsilon_2(\omega)$ data to infinite
isotopic masses, a procedure which would be desirable so as to
compare measured with calculated spectra. In the case of Ge,
however, the isotopic mass range available is much larger
$(^{70}$Ge$-^{76}$Ge). Moreover, one gains an additional factor of
2 (with respect to GaAs) because for Ge both atoms in the
primitive cell can be isotopically substituted. Nevertheless, such
an isotopic extrapolation procedure, which we believe to be
possible but difficult, has yet to be performed.

\begin{figure}[hbt]
\epsfxsize=.45\textwidth \centerline{\epsffile{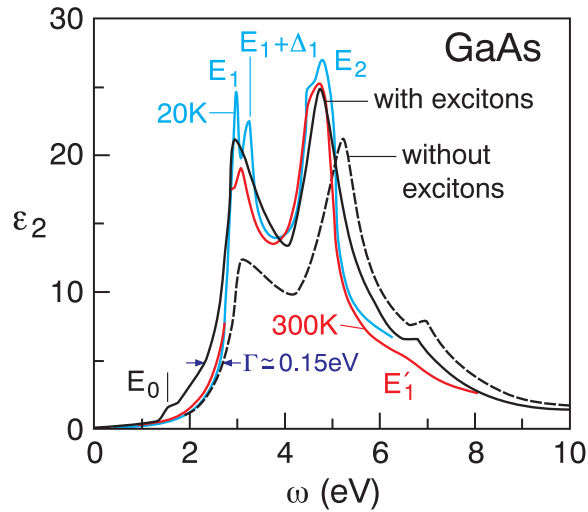}}
\caption[]{Spectra from Fig.~7 at 20K (blue) and 300K (red)
compared with the calculations of Rohlfing and Louie \cite{rohl}
with (thick solid line) and without (dashed line) excitonic
interactions but without spin-orbit coupling. The calculations
include an {\em ad hoc} Lorentzian broadening of 0.15 eV. Notice
the enhancement of the $E_1$ and $E_1+\Delta_1$ peaks induced by
the excitonic interaction.} \label{pssa8}
\end{figure}

We shall discuss here in some detail the temperature and isotope
effects of the $E_1$ and $E_1+\Delta_1$ features of Ge and Si.
They correspond to transitions between the top valence bands and
the bottom conduction band along the $\langle 111\rangle$
directions (labeled as $\Lambda$). Their splitting, 190 meV for
Ge, is due to spin-orbit interaction. Similar features are
observed for all materials of the germanium-zincblende family
\cite{yu68}: A red shift $(\Sigma_r)$ and a broadening with
increasing temperature.

\subsection{The $E_1$ and $E_1+\Delta_1$ critical points of
germanium: temperature and isotope effects}

R\"onnow {\it et al.} \cite{ronnow} were able to determine with a
remarkable accuracy the dependence of $\Sigma(T=0)$ on isotopic
mass for the $E_1$ and $E_1+\Delta_1$ CP's of germanium. We show
in Fig.~9 the third derivative of ellipsometrically determined
spectra of $\epsilon_1(\omega)$ and $\epsilon_2(\omega)$ for three
germanium wafers with different isotopic compositions measured at
30K. It is easy to see in this figure that the $E_1$  and $E_1 +
\Delta_1$ curves shift to larger frequencies with increasing
isotopic mass, in agreement with  the results found for the $E_g$
and $E_0$ critical points (Table I). The critical point parameters
$E_1, E_1+\Delta_1$, and $\Gamma$ (HWHM) obtained from the spectra
of Fig.~9 are plotted in Figs.~10 and 11 versus isotopic mass. In
spite of the considerable scatter of the experimental points, a
clear trend is apparent: $E_1, E_1+\Delta_1$, and even their
difference $\Delta_1$, increase with increasing mass. The solid
lines in Fig.~10 are fits to $E = E_{\infty} + BM^{\frac{1}{2}}$,
i.e., to the prediction of Eq. (8). These fits enable us to
determine the bare values of the critical points $E_{\infty}$;
they are listed in Table~II. Notice that the spin-orbit splitting
$\Delta_1$ shows a slight tendency to increase with increasing
mass, surely a result of the electron-phonon interaction;
quantitative calculations of this increase are not available.

\begin{figure}[hbt]
\epsfxsize=.45\textwidth
\begin{minipage}[b]{.35\textwidth}
{\epsffile{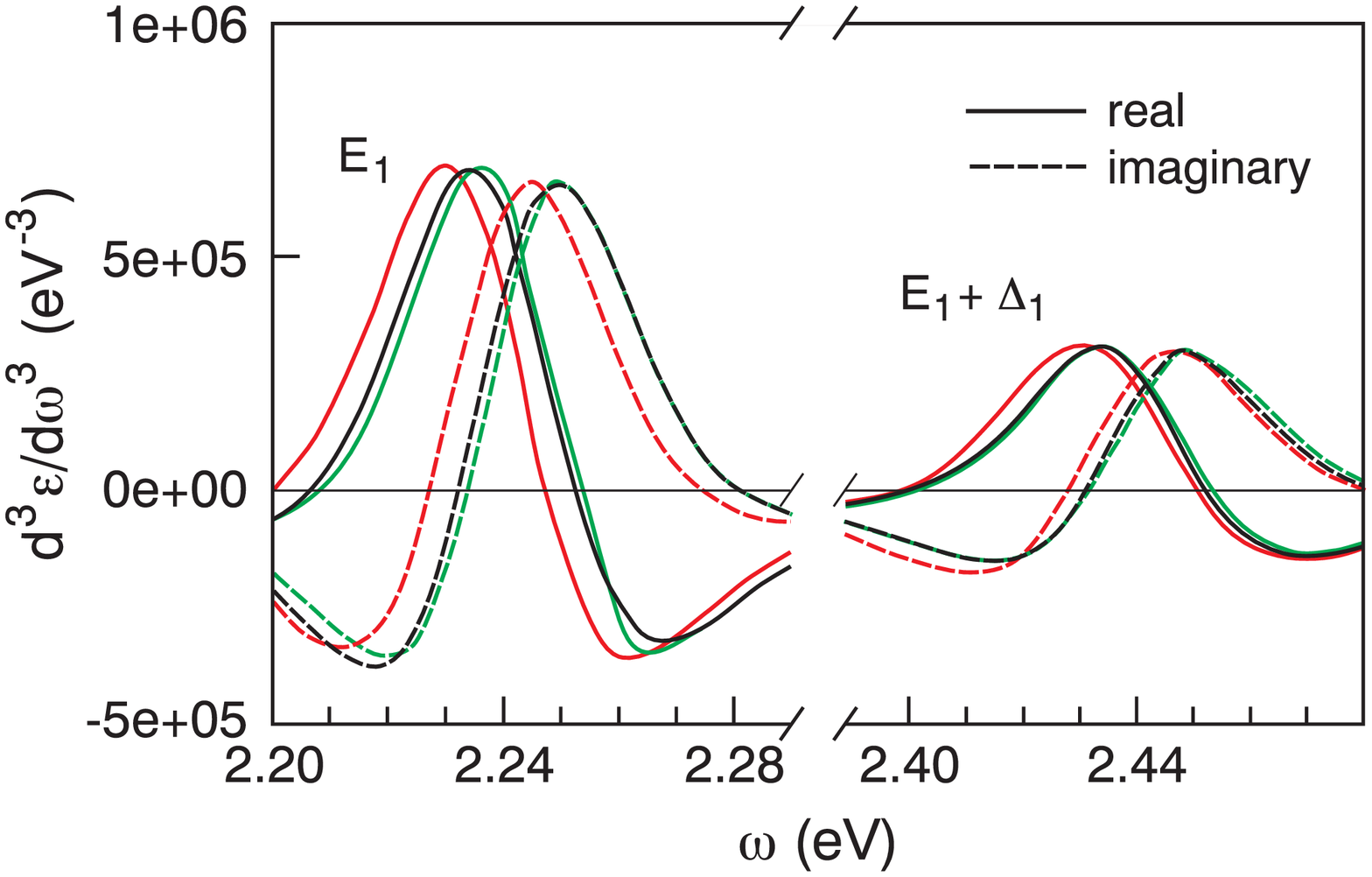}} \caption[]{Third derivative with respect to
photon energy of the dielectric function of natural Ge in the
vicinity of the $E_1$ and $E_1 + \Delta_1$ transitions. The solid
lines are fitted to the experimental data using analytical line
shapes. Red curves: $^{70}$Ge, black; natural Ge, green;
$^{75.6}$Ge, \cite{ronnow}.} \label{pssa9}
\end{minipage}\hfill
\epsfxsize=.45\textwidth
\begin{minipage}[b]{.45\textwidth}
{\epsffile{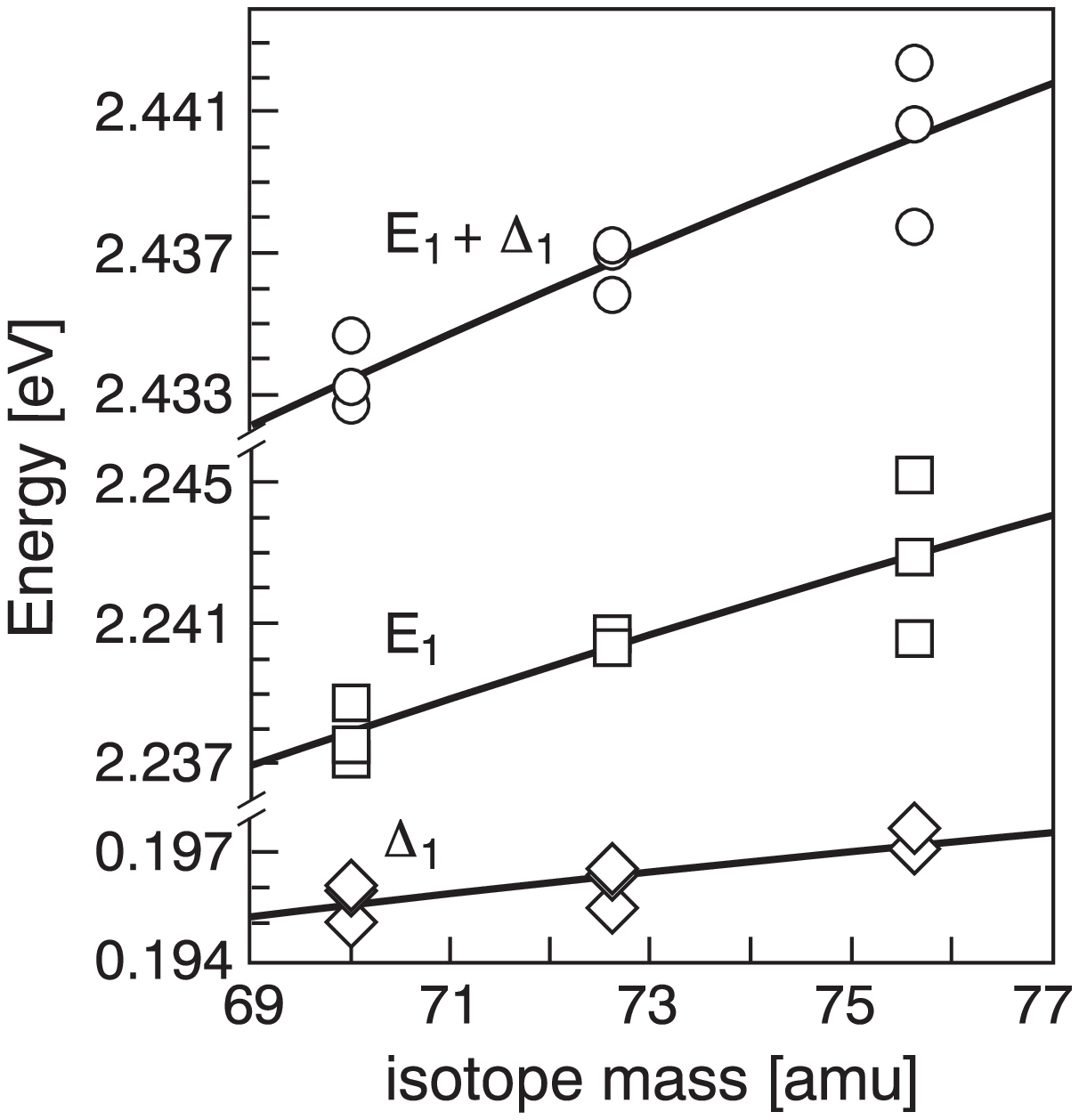}} \caption[]{ Energies of the $E_1$ and $E_1
+\Delta_1$ transitions in Ge, vs. isotope mass. Also shown is the
dependence of the spin-orbit splitting, $\Delta_1$, on isotope
mass. The solid lines are fits to $E = E_{\infty} +
BM^{-\frac{1}{2}}$ (see \cite{ronnow}).} \label{pssa10}
\end{minipage}
\end{figure}

Figure 11 displays the decrease in the Lorentzian HWHM parameters
of the $E_1$ and $E_1 + \Delta_1$ critical points of germanium
observed with increasing isotopic mass. The solid lines are fits
with the equation $\Gamma = \Gamma_{\infty} + BM^{-\frac{1}{2}}$,
based also on Eq.~(8), except that a small intercept
$\Gamma_{\infty}$ ($\backsimeq-6$ meV) allowed us to improve the
fit. This intercept, if meaningful, would represent a ``natural
width'' which would be present even when the atoms do not vibrate.
It could be due either to crystal imperfections, among them the
truncation at the surface, or, alternatively, to inadequacies in
the line shape used to fit the experimental spectra. In Ref.~71
logarithmic singularities were used (see Eq.~4 in \cite{ronnow})
which correspond to two-dimensional interband critical points.
Excitonic effects may contribute a small {\em negative}
$\Gamma_{\infty}$ to the fitted $\Gamma(M)$ as also may the
lowered dimensionality induced by the surface. The fitted value of
$B$ is $350\pm90$ meV/amu, in remarkably good agreement with band
structure based calculations (310 meV/amu \cite{zoll}).

\begin{figure}[hb]
\epsfxsize=.45\textwidth \centerline{\epsffile{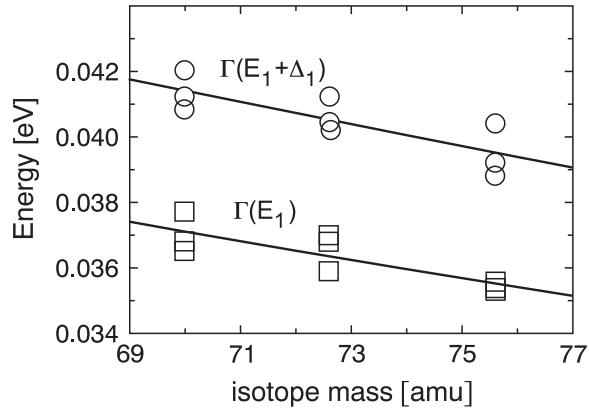}}
\caption[]{Lorentzian widths $\Gamma = -\Sigma_i$ of the $E_1$ and
$E_1 + \Delta$ transitions of germanium vs. isotopic mass. The
points are experimental. The lines represent fits to $\Gamma =
\Gamma_\infty + BM^{-\frac{1}{2}}$ \cite{ronnow}.} \label{pssa11}
\end{figure}

\subsection{The $E_1$ critical points of silicon: temperature and
isotope effects}

Work similar to that described in Sect.5.1 for germanium has been
performed for silicon by Lastras, {\it et al.} \cite{last26} using
$^{28}$Si, $^{30}$Si and natural silicon ($M = 28.09$). Because of
the small effects involved, the fits to the experimental
lineshapes were performed with an elaborate, highly accurate
Fourier transform technique \cite{asp}. The analysis of the data
is further complicated by the inability to resolve the spin-orbit
splitting $\Delta_1$ (for $S_i,\Delta_1\simeq$ 30 meV), and the
near-degeneracy of the $E_1$ CP's with the critical point $E_0'$.
Therefore, it was not possible to obtain data for the dependence
of $\Sigma_i$ on isotopic mass. The width of the $E_1$ transitions
measured directly at $T=10$ K is about 70 meV \cite{laut73}, a
value somewhat larger than that found for the other materials
under discussion. The value of $\Sigma_r(T=0)$ obtained from the
isotope shifts is $118\pm 22$ meV, somewhat larger than the
calculated value (71 meV) \cite{laut74} and much larger than the
value extracted from the measured temperature dependence of the
$E_1$ transition ($39\pm13$ meV) \cite{laut73}. The discrepancy
may be due to the complex nature of these transitions, as
mentioned above.

\begin{figure}[ht]
\epsfxsize=.40\textwidth \centerline{\epsffile{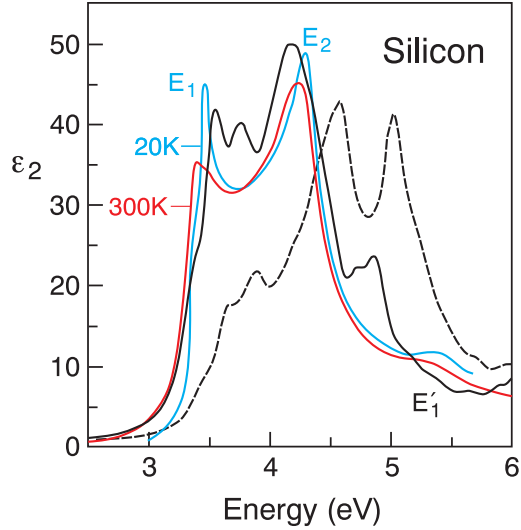}}
\caption[]{Ellipsometric spectra of $\epsilon_2(\omega)$ for
silicon \cite{albr} at two temperatures compared with the
calculations of Albrecht {\em et al.} \cite{alb} with (solid black
line) and without (dashed line) excitonic interaction. Note that
the peak calculated at 3.8 eV does not appear in the experimental
data and thus it is to be regarded as spurious, possibly due to
the coarse a {\bf k}-space interaction mesh.} \label{pssa12}
\end{figure}

We display in Fig.~12 the results of a recent {\em ab initio}
calculation of $\epsilon_2(\omega)$ for silicon, performed both
with (solid black line) and without (dashed line) excitonic
interaction \cite{alb}, compared with spectra measured
ellipsometrically at 20 and 300K \cite{laut73}. This figure
emphasizes the importance of comparing calculations performed for
static atoms with measurements at the lowest possible
temperatures: the height of the $E_1$ peak measured at 20K agrees
rather well with the calculated one. The calculated peak is,
however, broader than the measured one. This is a trivial effect
of having introduced in the calculation an artificial Lorentzian
broadening parameter of about 100 meV. The calculated $E_1$
structure is of obvious excitonic nature since it goes down
dramatically in strength if excitonic interaction is not included.
It peaks 100 meV higher than that measured at 20K. While 100 meV
is about the accuracy claimed for {\em ab initio} calculations, it
is tempting to speculate, in view of the results for $\Delta
E_1(T=0)$ given in Table~II, that the red shift of 100 meV found
in the experimental spectrum with respect to the calculated one is
due to the zero-point renormalization by the electron-phonon
interaction.

There has been some controversy concerning the peak exhibited by
the black solid curve of Fig.~12 at 3.8 eV. While in the original
publication \cite{alb} it was claimed that this peak corresponds
to a well-established experimental feature, Fig.~12 contradicts
this claim: It seems that the 3.8 eV feature is simply an artifact
of the coarse mesh used for the calculation \cite{card,albr}.

\section{Linear Optical Response Related to Surface States:
Effects of Electron-Phonon Interaction}

There have been in recent years a large number of investigations
of the response of electronic surface states to near-ir and
visible light. This response is rather weak and can best be
measured in a differential way which involves either subtracting
the responses to two inequivalently polarized light beams, or
letting the surface states disappear through passivation (e.g.,
oxidation \cite{cicca}). Even so, measurements in uhv are
difficult and time consuming. Consequently, only few experimental
investigations of the temperature dependence of the corresponding
spectra have been published \cite{cicca,olm,din}. These
investigations are mainly concerned with the gap between filled
and empty electronic surface states of cleaved [111] silicon
surfaces. For a critical discussion see Ref.~79.

The spectra of $2\times 1$ [111] silicon surfaces cleaved in
vacuum consist of a strongly polarized (for {\bf E}$\|~
[1\bar{1}0$], no response for {\bf E}$\|~[\bar{2}11$]) peak
centered at about 0.5 eV. This peak barely shifts with
temperature, a fact which has been attributed in Ref.~79 to a
compensation of the shifts induced by the various phonons
involved.\footnote{This is similar to the situation found for bulk
CuI \cite{ca,schweiz}.}

A red shift is, however, found for the low-frequency onset of this
peak (0.43 eV at 20 K, 0.34 eV at 300 K \cite{din}). The magnitude
of this shift is similar to that found for the lowest absorption
edge of many semiconductors (see Figs.~2 and 4). The onset of a
weak absorption peak, however, cannot be very reliably measured
and considerable differences exist among values reported for Si.
Nevertheless, several attempts have been made to interpret the red
shifts reported for the onset and the near lack of shift at the
peak \cite{olmstead}. The authors of these attempts basically
accept the latter fact and explain the shift of the onset as a
broadening of the peak structure.

An interesting observation concerning the response of the
$2\times1$ [111] silicon surface is the broadening with increasing
temperature which can, of course, be attributed to electron-phonon
interaction. We have seen in the previous sections that this
broadening is proportional to $T$ at high temperatures and becomes
$T$-independent for $T\rightarrow 0$, at least for bulk
semiconductors. In the case of surfaces one may argue that,
because of the reduced dimensionality, the coupling to the surface
electrons of individual surface phonons may be much stronger than
in the bulk and therefore the lowest order perturbation theory
which led to Eq.~6 may break down \cite{chen}. For the [111]
silicon surface there is no convincing evidence for this to be the
case.

It has been mentioned in Sect.~2.2 that, to lowest order in
perturbation theory, both $\Delta E(T)$ and $\Gamma(T)$ are
proportional to $\langle u^2\rangle$ ($u=$ phonon amplitude),
i.e., to $2\langle n_B\rangle +1$, where $\langle n_B\rangle$ is
an average Bose-Einstein factor (Eq.~(5)). This fact leads
immediately to the linear dependence on $T$ at high $T$ which has
been so profusely and fruitfully used throughout this article. the
surface spectroscopist will, however, have found in the literature
statements to the fact that the widths of spectral structures
induced by transitions between  electronic surface states are
proportional to $[2\langle n_B\rangle+1]^{1/2}$, i.e., to
$T^{1/2}$ at high $T$ \cite{cicca,olmstead}. This is a point where
communication between surface and bulk physicists could have been
better. The argument behind the linear-$u$ dependence emphasizes
the fact that the vibration, symmetric with respect to $u=0$,
slushes back and forth the frequency of the peak, thus producing a
broadening proportional to $u$. Those familiar with the
Debye-Waller theory will, however, realize that such a term
generates only a broad background, not a renormalized
band-structure preserving sharp critical points. The
renormalization of the band structure contains, in the spirit of
Fig.~1, only terms proportional to $2\langle n_B\rangle+1$.

\begin{figure}[hbt]
\epsfxsize=.35\textwidth \centerline{\epsffile{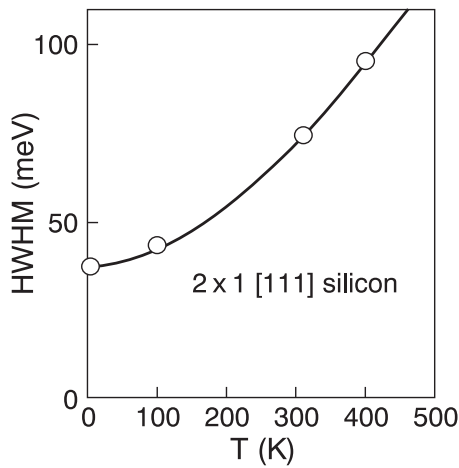}}
\caption[]{Widths~(HWHM) of the spectral peak observed around 0.5
eV for a $2\times1$-Si [111] surface at four temperatures. The
three points above 100 K are from \cite{cicca}, that near $T = 0$
was recently communicated to us by the authors of \cite{cicca}.
The line through the points represents a fit with the function
$[2n_B(\Theta/T)+1]\times\Gamma(0)$ which led to $\Theta =$ 340 K,
a rather reasonable value for silicon.} \label{psssa13}
\end{figure}

Unfortunately, the only measurements vs. temperature of the width
of the $2\times 1$ Si [111] peak were reported solely for three
temperatures \cite{cicca} (see Fig.~13; a fourth point, close to
$T=0$, has been recently communicated to us by the authors of
\cite{cicca}) and they hardly possess enough information to refute
experimentally the square-root hypothesis. In fact, they were
presented in \cite{cicca} as supporting it. We have plotted in
Fig.~13 these points vs. $T$ and fitted them with a
single-oscillator $2n_B+1$ expression. The fit for an average
frequency $\theta =$ 340K, is quite satisfactory, slightly better
than the square-root fit of Fig.~2 of \cite{cicca}.

\section{Dependence of Other Physical Properties on Temperature
and Isotopic Mass}

The Bose-Einstein fits and measured temperature dependences of a
wide variety of physical properties of semiconductors, in
connection with linear asymptotic expansions, can also be used to
estimate the zero-point renormalization of these properties and
the corresponding isotope effects. In this section we briefly
discuss the long-wavelength refractive index, the phonon
frequencies and the elastic constants for elemental
semiconductors. The semiempirical expressions used for the fits
are basically the same as those discussed in previous sections for
interband transition energies. However, the $M^{-\frac{1}{2}}$
dependence of the zero-point renormalizations, given in Eq.~(8),
must be modified depending on the type of property under
consideration. This dependence still holds for the long wavelength
refractive index since it is directly related to
$\epsilon_1(\omega)$ and $\epsilon_2(\omega)$ (a straightforward
calculation suffices to prove it).

In the case of phonon properties, such as phonon frequencies, the
$M^{-\frac{1}{2}}$ dependence must be replaced by $M^{-1}$. This
is easily understood.  The diagram, corresponding to ($ii$), which
represents an anharmonic renormalization of a phonon frequency, is
obtained by replacing the horizontal electron propagators by curly
phonon propagators. Hence, four phonon amplitudes come together at
the vertex, giving a matrix element proportional to $u^4$, i.e.,
to $M^{-1}$. For the elastic constants, proportionality to
$M^{-{\frac{1}{2}}}$ obtains.

\subsection{Long wavelength refractive index: temperature
dependence and isotope effects}

The temperature dependence of the long wavelength refractive index
plays an important role in the tuning of optical and
optoelectronic  devices. Early theoretical work was concerned only
with the temperature region in which the $T$-dependence is linear
\cite{yu,wehner}. Recently, {\em ab initio} calculations over a
broad temperature range, including the zero-temperature
renormalization, have been performed \cite{karch}. Experimental
results are given in \cite{ruf} for diamond and in \cite{karch}
for Si and Ge.

The following fitted expression is given in Ref.~82 for the long
wavelength refractive index $n(T)$ of diamond.

\begin{equation} 
n(T) ~ = ~ n_0 + ~\frac{A}{2}~ \Big[ 2 n_B ~(\Omega,T)~ +1 \Big]
\end{equation}

\noindent where $\Omega = 711$ cm$^{-1}$, A = 0.19 and $n_0 =$
2.377.  This expression implies that  the zero-point
renormalization is $\Delta n(T=0)=$ 0.085, rather close to the
value 0.07 obtained by linear extrapolation from Fig.~5 of Ref.~7.
Note, however, that the value of $\Delta n$ given in Table I of
Ref.~7 is three times larger and thus must be erroneous, since
inconsistent with Fig.~5 of that reference.

From the zero-point renormalization $\Delta n(T=0)=+0.085$, the
dependence on isotopic mass:

\begin{equation} 
\left(\frac{\partial n_0}{\partial M}\right) ~=~
4\times10^{-4}~K^{-1}
\end{equation}

\noindent is obtained. It is slightly larger than the value
$3\times10^{-4}$ K$^{-1}$ calculated in Ref.~7. Fabry-Perot
interference fringes are a very accurate method to measure
differences in refractive indices. They could be used to measure
the coefficient of Eq.~(14) by comparing two plane-parallel
diamond wafers made out of $^{12}$C and $^{13}$C, respectively.
The expected value of this coefficient, however, is so small that
differences in the samples resulting from their fabrication, but
due to reasons other than their isotopic mass, would be likely to
obliterate the expected results. At this time it would also be
very difficult to measure the corresponding derivatives of $n_0$
for Ge and Si which, nevertheless, can be easily obtained from the
calculations in \cite{karch}.

\subsection{Anharmonic self-energy of phonons: effects of
temperature and isotopic substitution}

Although anharmonic self-energies can be found for all phonons in
the dispersion relations, most of the information available
concerns the Raman phonons, i.e., the optical phonons for {\bf q}
= 0. In most cases analyzed in detail, the equivalent of the two
terms in Fig.~1 (with the electron lines replaced by  phonon
propagators) suffice to explain the measured $\Sigma(\Omega)$
\cite{deb,widu,widulle}.

Let us first discuss the dependence of the phonon linewidth (HWHM
= $-\Sigma_i$) of the equivalent of Fig.~1 for phonons. No
contribution of the phonon-phonon equivalent of ($ii$) exists
\cite{menen,deb65}. The contribution of ($i$) to $\Sigma_i$
represents a decay of the phonon under consideration (frequency
$\Omega$, {\bf q}=0 for the Raman phonon) into two phonons, with
frequency and wavevector conservation ($\Omega=\Omega_1+\Omega_2$;
$\rm{\bf q}={\bf q}_1+{\bf q}_2$). In the case of Ge and Si
excellent fits to the measured temperature dependence of the HWHM
are obtained with the expression \cite{menen,deb65}:

\begin{equation} 
\Gamma(T) = -\Sigma_i(T) = \Gamma_0~ \Big[1+ n_B(\Omega_1,T)
+n_B(\Omega_2,T)\Big]
\end{equation}

\noindent where $\Omega_1 = 2\Omega_2 = 0.66\Omega$ ($\Omega_1$
and $\Omega_2$ correspond to LA and TA phonons at the edge of the
BZ), and the zero-point renormalization $\Gamma_0\backsimeq 0.5$
cm$^{-1}$, which has indeed been shown to be proportional to
$M^{-1}$.\footnote{{\em See Fig.~4 in Ref.~83, where the
contribution of isotopic disorder to $\Gamma$ has also been
included. This contribution amounts to 0.035 cm$^{-1}$ HWHM for a
mixture of equal parts of $^{28}$Si and $^{30}$Si.}} The energy
conservation constraint $\Omega=\Omega_1+\Omega_2$ is only
required for the evaluation of $\Sigma_i$.

The frequency shift $\Sigma_r$, including the contribution
equivalent to ($ii$), is obtained by summing over all possible
{\em virtual} intermediate states, with ($\rm{\bf q}={\bf
q}_1+{\bf q}_2$) but without the requirement of energy
conservation. However, because of the energy denominators which
appear in the second-order perturbation expression corresponding
to $\Sigma_r$, two phonons with $\Omega_1 + \Omega_2$ near
$\Omega$ often give the dominant contributions to $\Sigma_r$.
Hence, it is common to find in the literature fits to the measured
$\Sigma_r(T)$ with Eq.~(15) and the restriction
$\Omega=\Omega_1+\Omega_2$. Such a fit is found for silicon in
Fig.~3 of Ref.~83. It leads to an estimate $\Sigma_r= -5.6$
cm$^{-1}$ for the zero-point anharmonic renormalization of the
phonon frequency of Si (bare frequency at $T=0$  equal to 531
cm$^{-1}$). This zero-point renormalization, proportional to
$M^{-1}$, decreases in absolute value by 0.39 cm$^{-1}$ when
replacing $^{28}$Si by $^{30}$Si.

\subsection{Elastic Constants}

The temperature dependence of the elastic stiffness moduli
C$_{11}(T)$, C$_{12}(T)$ and C$_{44}(T)$ has been measured for
diamond \cite{mcskim}, silicon \cite{mcskimin}, and germanium
\cite{mcskimin}. In spite of the temperature range, which was
limited to $T$ slightly below the Debye temperature, it has been
possible to estimate the position of the linear asymptote required
for the determination of the zero-point renormalization. As
already mentioned, this renormalization is related to that of the
long-wavelength acoustic phonons and turns out to be proportional
to $M^{-{\frac{1}{2}}}$.

In Table~I of [49] the values of the elastic constants, the
estimated temperature renormalization and their logarithmic
derivatives with respect to $M$, calculated by using the
$M^{-\frac{1}{2}}$ dependence on mass, are displayed. No
experimental data are available for these derivatives but, in
spite of the small values involved, they may be measurable using
the highly accurate ultrasonic propagation technique.

It is interesting to mention that a value of $\partial
 lnC/\partial M = 5\times10 ^{-3}$ was reported in \cite{ram} for
diamond. A point was made in that paper of the fact that
correspondingly, $^{13}$C diamond must be significantly harder
than $^{12}$C diamond. This $\partial lnC/\partial M$ is about an
order of magnitude larger than the value obtained from Table~I of
Ref.~[49], which implies insignificantly harder $^{13}$C diamond.
In a following paper \cite{vogel} an error in the interpretation
of the experiments of \cite{ram} was corrected. This correction,
and an estimate with a theoretical model, led to a value of
$\partial lnC/\partial M \leq 10^{-3}$, in agreement with the
value given in  Table~I $(3.5\times 10^{-4})$.

\section{Conclusions}

We have discussed the renormalization of a broad range of physical
properties of tetrahedral semiconductors by the electron-phonon
interaction. These properties include the dielectric functions and
the corresponding electronic transition energies (both bulk and
surface contributions), the lattice parameter, the phonon
frequencies and the elastic stiffness constants. Particular
attention has been paid to the zero-point $(T=0)$
renormalizations, which can be changed by changing the isotopic
masses. Semiempirical expressions representing these
renormalizations vs. temperature have been given. Especially
powerful for the interpretation of experimental results has been
shown to be the determination of the asymptotic linear behavior
vs. temperature at high temperature, either with the naked eye or,
even better, from a fit to a well-founded analytical expression.

In this article we have only discussed elemental and binary
semiconductors. A few cases involving materials with three
different elements or more can be found in the literature. Of
particular interest are the I-III-VI chalcopyrites containing as
element either copper or silver (e.g., CuGaSe$_2$ \cite{quin},
which is isoelectronic to ZnSe; AgInSe$_2$ \cite{aliy}
isoelectronic to CdSe; AgGaSe$_2$ \cite{choi}). These materials
all exhibit anomalies in the temperature dependences of the gap
which are reminiscent of those observed for the copper halides: at
low $T$ the gaps increase with increasing $T$, whereas they
decrease at higher $T's$. While it is easy to conjecture that the
anomalies are due to the presence of 3d electrons of Cu (or 4d of
Ag) in the valence bands, measurements of the dependence of the
low temperature gap on isotopic mass would be desirable in order
to tune these anomalies. Many of these chalcopyrites (e.g.,
Cu,Ga,Se) have three elements with more than one stable isotope
each.

The present author was delighted to see, after completing this
paper, a very recent article \cite{vis} in which the temperature
renormalization of eight prominent structures, seen  from 2 to 5
eV, in the reflectance anisotropy spectra of (100) InP surfaces
were reported. The experimental data for the temperature shifts
were fitted with single oscillators, i.e., with expressions
similar to Eq.(6). The zero-point renormalizations of most of the
eight observed structures lie around -50 meV, rather similar to
those of the bulk CP's. The fitted oscillator frequencies are
close to the average phonon frequencies of bulk InP. As required
by the oscillator fit, all observed frequencies depend linearly on
$T$ at high $T$. A square-root dependence is thus ruled out, at
least for the frequency shifts.

\clearpage
\newpage
\newcommand{\rb}[1]{ \raisebox{1.5ex}[-1.5ex]{#1}}
\renewcommand{\arraystretch}{1.5}
\begin{table}
\begin{tabular}{|l|r r|r@{.}l r@{.}l|r@{.}l r@{.}l| } \hline%
& \multicolumn{2}{|c|}{$\Delta E(T=0)$} &
 \multicolumn{4}{|c|}{$dE/dM$}& %
 \multicolumn{2}{|c|}{$\Delta E_{th}(T=0)$} & %
 \multicolumn{2}{|c|}{$dE_{th}/dM$} %
\\ \hline
Ge($E_g$)& -53 & -53 & 0&36&  0&37 &
 \multicolumn{2}{l|}{-13} & 0&09
\\ \hline %
Ge($E_0$)     & -60 & -71  & 0&49 & 0&41 & \multicolumn{2}{l|}{-34}  & 0&24 \\ \hline%
Si($E_g$)     & -64 & -50  & 0&86 & 1&10 & \multicolumn{2}{l|}{+8}   & -0&14 \\ \hline%
diamond($E_g$)& -615& -340 & 13&6 & 17&1
&\multicolumn{2}{l|}{-88}  &  3&6 \\ \hline %
&&& \multicolumn{2}{c}{} & \multicolumn{2}{c|}{}
&\multicolumn{2}{c|}{} & 0&16 \\%
\rb{GaAs($E_0$)} &\rb{-90} & \rb{-110} &
\multicolumn{2}{l}{\rb{\hphantom{1}0.39}} &
\multicolumn{2}{l|}{\rb{\hphantom{1}0.32}} &
\multicolumn{2}{l|}{\rb{-38}} & 0&11
\\ \hline
GaP($E_g)$ &-120 && \multicolumn{2}{c}{} & \multicolumn{2}{c|}{}
& +5&3 & \multicolumn{2}{|c|}{}   \\ \hline%
&&& 0&22 & 0&19 & \multicolumn{2}{c|}{} & 0&12 \\ %
\rb{ZnSe($E_0$)} & \rb{-65} & \rb{-64} & 0&22 & 0&26 &
\multicolumn{2}{l|}{\rb{-15}}  & 0&04 \\ \hline %
&&& -0&08 & \multicolumn{2}{c|}{} & \multicolumn{2}{c|}{} &
\multicolumn{2}{c|}{} \\ \rb{CuCl($E_0$)} & \rb{+30} &   & +0&36
&\multicolumn{2}{c|}{}
 & \multicolumn{2}{l|}{\rb{\hphantom{1}-2.8}}
 & \multicolumn{2}{c|}{} \\ \hline%
&&& -0&11 & \multicolumn{2}{c|}{} & \multicolumn{2}{c|}{}
& \multicolumn{2}{c|}{} \\%
\rb{CuBr($E_0$)} & \rb{+6}  &   & +0&12 & \multicolumn{2}{c|}{} &
\multicolumn{2}{l|}{\rb{\hphantom{1}-1.4}}
& \multicolumn{2}{c|}{} \\ \hline%
& & & -0&55 & \multicolumn{2}{c|}{} & \multicolumn{2}{c|}{}
&\multicolumn{2}{c|}{} \\ \rb{CuI($E_0$)} & \rb{+5} & &
\multicolumn{2}{c}{--} & \multicolumn{2}{c|}{}
& \multicolumn{2}{l|}{\rb{\hphantom{1}-3.5}} & \multicolumn{2}{c|}{} \\ \hline%
\end{tabular}

\caption[]{Zero-point anharmonic renormalization $\Delta E(T=0)$
of the lowest gaps of several zincblende-type semiconductors (in
meV) together with its derivative with respect to the isotopic
mass (in meV/amu). Direct and indirect gaps are labeled $E_0$ and
$E_g$, respectively. For completeness, the thermal expansion
contribution to the zero-point parameters is also listed $(\Delta
E_{th}$ and $dE_{th}/dM$). For the binary compounds the
derivatives with respect to cation (above) and anion (below)
masses are given. The first number under $\Delta E(T=0)$ has been
measured through linear extrapolation, except for diamond where a
value calculated in \cite{zoll} is given. The first number under
$dE/dM$ has been measured through isotopic substitution. The
second numbers under $\Delta E$ and $dE/dM$ were estimated from
the measured values of $dE/dM$ and $\Delta E$, respectively, using
Eq.(8). For sources see text.}
\end{table}

\newpage
\renewcommand{\arraystretch}{1.5}
\begin{table}
\begin{tabular}{|l|c|c|c|c|}
\hline%
& $dE_1/dM$ & $d(E_1+\Delta_1)/dM$ & $\Delta E_1(T=0)$ &
$\Delta (E_1\Delta_1)(T=0)$   \\ \hline %
& 0.9~~  & 1.2       & 131   & \\ %
\rb{Ge}  & ~0.45$^a$ &       & ~~65$^a$ & \rb{175} \\ \hline %
& 2.0~~  &           & 118   & \\%
\rb{Si} & ~1.2$^a$~~ & \rb{--} &~~71$^a$ & \rb{--} \\ \hline
\end{tabular}

\vspace{0.5cm} \noindent $^a$calculated \cite{zoll,laut}

\caption[]{Derivatives of the $E_1$ and $E_1+\Delta_1$ critical
point energies of germanium at low temperatures as obtained from
Figs.~(9) and (10). Also, corresponding zero-point gap
renormalizations estimated from the above values with Eq.(8) (in
meV/amu and meV) and corresponding values for the $E_1$ critical
point of silicon \cite{last26}. The experimental values measured
by isotopic substitution are nearly a factor of two larger than
those calculated using empirical pseudopotentials
\cite{zoll,laut}.}
\end{table}

\newpage

\begin{thebibliography}{94}
\bibitem{einstein} A. Einstein, Ann.~Physik {\bf 22}, 180 (1907).
\bibitem{nernst} W.~Nernst and Lindemann, Preus. Akad. Wiss., Phys.
                   Math. K., (1911), p. 494.
\bibitem{debye} P.~Debye, Ann. Physik {\bf 39}, 789 (1912).
\bibitem{grun} E.~Gr\"uneisen, Handbuch der Physik {\bf 10}, 22
                (1926).
\bibitem{dong} J.~Dong, O.F.~Shankey, and C.W. Myles, Phys. Rev. Lett.
               {\bf 86}, 17 (2001).
\bibitem{einst} A.~Einstein, Ann. Physik {\bf 35}, 79 (1911).
\bibitem{karch} K.~Karch, T. Dietrich, W. Windl, P. Pavone, A.P.
                  Mayer, and D.~Strauch, Phys. Rev. B {\bf 53}, 7259 (1996).
\bibitem{fan} H.Y.~Fan, Phys, Rev. {\bf 82}, 900 (1951).
\bibitem{allen9} P.B.~Allen and M. Cardona, Phys. Rev. B {\bf 27}, 4760
(1983); M.L. Cohen and D.J. Chadi, in ``Handbook of
Semiconductors'', Vol. II, M. Balkanski, editor, (North Holland,
Amsterdam, 150) p. 155.
\bibitem{deb} A.~Debernardi, Solid State Commun. {\bf 113}, 1 (2000).
\bibitem{king} D.~King-Smith, R.J.~Needs, V.~Heine, and
                M.J.~Hodgson, Europhys. Lett. {\bf 10}, 569 (1989).
\bibitem{rohl} M.~Rohlfing and S.G.~Louie, Phys. Rev. Lett. {\bf
               81}, 2312 (1998).
\bibitem{bene} L.X.~Benedict, E.L.~Shirley, and R.B.~Bohm, Phys. Rev. Lett.
               {\bf 80}, 4514 (1998).
\bibitem{alb} S.~Albrecht, L.~Reining, R.~del~Sole, and G.~Onida,
               Phys. Rev. Lett. {\bf 80}, 4510 (1998).
\bibitem{brust} J.D.~Brust, J.C.~Phillips, and F.~Bassani,
                 Phys. Rev. Lett. {\bf 9}, 95 (1962).
\bibitem{cohen} M.L.~Cohen and J.~Chelikowsky, Electronic
                Structure and Optical Profperties of Semiconductors, Springer,
                 Heidelberg, 1989.
\bibitem{etch17} P.~Etchegoin, J. Kircher, and M. Cardona,
               Phys. Rev. B {\bf 47}, 10292 (1993); {\em ibid}~
               Phys. Rev. B {\bf 46}, 15139 (1992).
\bibitem{ronn} D.~R\"onnow, L.F. Lastras-Mart\'{\i}nez, M. Cardona, and P.V. Santos,
                   J.~Opt.~Soc.~Am.~A {\bf 16}, 568 (1999).
\bibitem{last} A.~Lastras-Mart\'{\i}nez, R.E.~Banderas-Navarro, and L.F.~Lastras-Mart\'{\i}nez,
               Thin Solid Films {\bf 373}, 207 (2000).
\bibitem{ali} C.~Alibert, A.M.~Joullie, A.~Joullie, and R.~Ranvaud,
              Nuovo Cimento {\bf 39}, 427 (1977).
\bibitem{ca21} M.~Cardona, M.H.~Grimsditch, D. Olego in Light
               Scattering in Solids, J.L. Birman and H.Z. Cummins (Eds.), Plenum
               Publ. Co., New York (1979), p. 249.
\bibitem{etch22} P.~Etchegoin, J.~Kircher, M.~Cardona, and
C.~Grein, Phys.~Rev.~B {\bf 45}, 11721 (1992).
\bibitem{theo} G.~Theodorou, phys. stat. sol. b {\bf 211}, 847
(1999); {\em ibid} {\bf 211}, 29 (1999).
\bibitem{hughes} J.L.R. Hughes and J.E. Sipe, Phys. Rev. B {\bf
53}, 10751 (1998).
\bibitem{card} M.~Cardona, L.F.~Lastras-Mart\'{\i}nez, and D.E.~Aspnes,
               Phys. Rev. Lett. {\bf 83}, 3970 (1999).
\bibitem{last26} L.F.~Lastras-Mart\'{\i}nez, T. Ruf, M. Konuma, M.~Cardona,
               and D.E.~Aspnes, Phys. Rev. B {\bf 61}, 12946 (2000).
\bibitem{coll} A.T.~Collins, S.C. Lawson, G.~Davies, and H.~
               Kanda, Phys. Rev. Lett. {\bf 65}, 891 (1960).
\bibitem{etch} P.~Etchegoin, J.~Weber, M.~Cardona, W.L.~Hansen,
               K.~Itoh, and E.E.~Haller, Solid State Commun.
                   {\bf 83}, 843 (1992).
\bibitem{allen29} P.B.~Allen, Phil. Mag. {\bf B70}, 527 (1994).
\bibitem{ant} E.~Anton\v{c}ik, Czech. J. Phys. {\bf 5}, 449 (1955).
\bibitem{cicca} F.~Ciccacci, S. Selci, G. Chiarotti, and P. Chiaradia,
               Phys. Rev. Lett. {\bf 56}, 2411 (1986).
\bibitem{pavone} P.~Pavone and S.~Baroni, Solid State Commun.
               {\bf 90}, 295 (1994).
\bibitem{deber} A.~Debernardi and M.~Cardona, Phys. Rev. B
              {\bf 54}, 11305 (1996).
\bibitem{london} H.~London, Z. Phys. Chem. {\bf 16}, 302 (1958).
\bibitem{busch} R.C.~Buschert, A.E.~Merlini, S.~Pace,
               S.~Rodr\'{\i}guez, and M.H.~Grimsditsch, Phys. Rev. B {\bf 38},
               5219 (1988).
\bibitem{sozont} E.~Sozontov, L.X.~Cao, A.~Kazimirov, V.~Kohn,
                 M.~Konuma, M.~Cardona, and J.~Zegenhagen, Phys. Rev. Lett. {\bf 86},
                 5329 (2001).
\bibitem{reeber} R.R. Reeber and K. Wang, J. Electronic Materials
{\bf 25}, 63 (1996).
\bibitem{wein} B.A.~Weinstein and R.~Zallen, in ``Light Scattering
                in Solids IV'', M.~Cardona and G.~G\"untherodt, editors,
                (Springer, Heidelberg, 1984) p. 463.
\bibitem{serrano} J.~Serrano, unpublished.
\bibitem{lyon} K.G.~Lyon, G.L.~Salinger, C.A.~Swenson, and
 K.G.~White, J. Appl. Phys. {\bf 48}, 865 (1977).
\bibitem{allen41} P.B.~Allen, Phys. Rev. B {\bf 18}, 5217 (1978).
\bibitem{karai} D.~Karaiskaj, M.L.W.~Thewalt, T.~Ruf, M.~Cardona,
               H.J.~Pohl, G.G.~Deviatych, \linebreak
                P.G.~Sennikov, and H.~Riemann, Phys.
               Rev. Lett. {\bf 86}, 6010 (2001).
\bibitem{quint} M.~Quintero, C.~Rincon, R.~Tovar, and J.C.~Woolley,
                J. Phys. Condens. Matter {\bf 4}, 1281 (1992).
\bibitem{zoll} S.~Zollner, M.~Cardona, and S.~Gopalan, Phys. Rev. B
              {\bf 45}, 3376 (1992).
\bibitem{yu} P.Y.~Yu and M. Cardona, Phys. Rev. B {\bf 2}, 3193
               (1998).
\bibitem{laut} P.~Lautenschlager, P.B.~Allen, and M~ Cardona, Phys.
                Rev. B {\bf 31}, 2163 (1985).
\bibitem{thur} C.D.~Thurmond, J. Electrochem. Soc. {\bf 122}, 1133
               (1975).
\bibitem{parks} C.~Parks, A.K.~Ramdas, S.~Rodr\'{\i}guez, K.M.~Itoh,
               and E.E.~Haller, Phys. Rev. B {\bf 49}, 14244 (1994).

\bibitem{temp} The temperature dependence of $a_0$ for silicon and
                a fit similar to that of Fig.~3 will be found in: M. Cardona,
                 Festschrift in honor of F. Bassani (Scuola Normale Superiore,
                 Pisa, 2001).
\bibitem{hollow} H.~Holloway, K.C.~Hass, M.A.~Tamar, T.R.~Anthony,
                  and W.F.~Banholzer, Phys. Rev. B {\bf 44}, 7123 (1991).
\bibitem{gob} A.~G\"obel, T.~Ruf, M.~Cardona, C.T.~Lin,
              J.~Wrzesinski, M.~Steube, K.~Reimann, J.-C.~Merle, and M.~Joucla,
              Phys. Rev. B {\bf 57}, 15183 (1998).
\bibitem{garro} N.~Garro, A.~Cantarero, M.~Cardona, A.~G\"obel,
              T.~Ruf, and K.~Eberl, Phys. Rev. B {\bf 54}, 4732 (1996).
\bibitem{varsh} Y.P.~Varshni, Physica {\bf 34}, 149 (1967).
\bibitem{vina} L.~Vi\~{n}a, S.~Logothetidis, and M. Cardona, Phys.
                 Rev. B {\bf 30}, 1979 (1984).
\bibitem{mano} A.~Manoogian and A. Leclerc, Can. J. Phys. {\bf
                       57}, 1766 (1979).
\bibitem{pass} R.~P\"assler, Solid State Electr. {\bf 39}, 1311 (1996).
\bibitem{collins} A.T.~Collins S.C.~Lawson, G.~Davies, and H.~Kanda,
               Phys. Rev. Lett. {\bf 65}, 891 (1990).
\bibitem{ca58} M.~Cardona and N.E.~Christensen, Solid State
               Commun. {\bf 58}, 421 (1986).
\bibitem{passl} R.~P\"assler, phys. stat. sol.(b), {\bf 200}, 155
               (1997).
\bibitem{gobel} A.~G\"obel, T.~Ruf, J.M.~Zhang, R.~Lauck, and M.
                 Cardona, Phys. Rev. B {\bf 59}, 2749 (1999).

\bibitem{zung} A.~Zunger and M.L.~Cohen, Phys. Rev. B {\bf 18},
                 5449 (1978); {\it ibid} {\bf 20}, 4082 (1979).
\bibitem{yin14} M.T.~Yin and M.L.~Cohen, Solid State Commun. {\bf
               43}, 391 (1982).
\bibitem{yin15} M.T.~Yin and M.L.~Cohen, Phys. Rev. B {\bf 26},
 3259 (1982).
\bibitem{godby} R.W.~Godby, M. Schl\"uter, and L.J~Sham, Phys.
               Rev. Lett. {\bf 56}, 2415 (1986).
\bibitem{ca} M.~Cardona, Phys. Rev. {\bf 129}, 69 (1963).
\bibitem{schweiz} J. Serrano and C.~Schweitzer, unpublished
                    results.
\bibitem{lauten} P. Lautenschlager, M. Garriga, S. Logothetidis, and M. Cardona,
                 Phys. Rev. B {\bf 35}, 9174 (1987).
\bibitem{yu68}P.Y.~Yu and M. Cardona, ``Fundamentals of
Semiconductors'', (Springer, Heidelberg, 2001, 3rd edition), p.
320.
\bibitem{law} P.~Lawaetz, Thesis, The Technical University of
                 Denmark, Lyngby, 1978 (unpublished).

\bibitem{aspnes} D.E. Aspnes and A.A. Studna, Phys. Rev. B {\bf 7},
               4605 (1973).
\bibitem{ronnow} D. R\"onnow, L.F.~Lastras-Mart\'{\i}nez, and
               M.~Cardona, Eur. Phys. J. {\bf B5}, 29 (1998).
\bibitem{asp} D.E.~Aspnes and H.~Arwin, J.O.S.A. {\bf 73}, 1759
              (1983).
\bibitem{laut73} P.~Lautenschlager, M.~Garriga, L.~Vi\~{n}a, and M.~Cardona,
                  Phys. Rev. B {\bf 36}, 4821 (1987).
\bibitem{laut74} P.~Lautenschlager, P.B.~Allen, M.~Cardona, Phys.
                   Rev. B {\bf 33}, 5501 (1986).
\bibitem{albr} M. Cardona, L.F. Lastras-Mart\'{\i}nez, and D.E. Aspnes,
               Phys. Rev. Lett. {\bf 83}, 3970 (1999); \linebreak
               S. Albrecht, L. Reining, G. Onida, V. Olevano, and
               R.~DelSole, Phys. Rev. Lett. {\bf 83}, 1971 (1999).

\bibitem{selci} S.~Selci, P. Chiaradia, F. Ciccacci, A. Cricenti,
                N. Sparvieri, and G. Chiarotti, Phys. Rev. Lett. {\bf 31}, 4096
               (1985).
\bibitem{olm} M.J.~Olmstead and N.M.~Amer, Phys. Rev. B {\bf 33}, 2564
              (1986).
\bibitem{din} N.J.~DiNardo, J.E.~Demuth, W.A.~Thompson, and P.
              Avouris, Phys. Rev. B {\bf 31}, 4077 (1985).
\bibitem{olmstead} M.J.~Olmstead and D.J.~Chadi, Phys. Rev. B {\bf
                  33}, 8402 (1986).
\bibitem{chen} C.D.~Chen, A. Selloni, and E. Tosatti, Phys. Rev. B
               {\bf 30}, 7067 (1984).

\bibitem{wehner} R.K.~Wehner and R.~Klein, Physica {\bf 62}, 161
                 (1972).
\bibitem{ruf} T.~Ruf, M.~Cardona, C.S.J.~Pickles, and R.~Sussmann,
              Phys. Rev. B {\bf 62}, 16578 (2000).

\bibitem{widu} F.~Widulle, T.~Ruf, M.~Konuma, I.~Silier,
                W.~Kriegseis, V.I.~Ozhogin, and M.~Cardona,\\ Solid State Commun.
                {\bf118}, 1 (2001).
\bibitem{widulle} F.~Widulle, J. Serrano, and M.~Cardona,
                  submitted to Phys. Rev. B.
\bibitem{menen} J.~Men\'endez and M.~Cardona, Phys. Rev. B {\bf 29},
                 2051 (1984).
\bibitem{deb65} A.~Debernardi, S. Baroni, and E.~Molinari, Phys.
                    Rev. Lett. {\bf 75}, 1819 (1995).
\bibitem{mcskim} H.J.~McSkimin and P.~Andreatch, J. Appl. Phys.
               {\bf 43}, 2944 (1972).
\bibitem{mcskimin} H.J.~McSkimin, J. Appl. Phys. {\bf 24}, 988
               (1988).
\bibitem{ram} A.K.~Ramdas, S.~Rodr\'{\i}guez, M. Grimsditch,
              T.R.~Anthony, and W.F.~Banholzer, Phys. Rev. Lett. {\bf 71}, 189
              (1993).
\bibitem{vogel} R.~Vogelgesang, A.K.~Ramdas, S.~Rodr\'{\i}guez,
               M.~Grimsditch, and T.R.~Anthony, Phys. Rev. B {\bf 54}, 3989
               (1996).
\bibitem{quin} M.~Quintero, C. ~Rincon, R.~Tovar, and J.C.~Woolley, J.
               Phys. Cond. Mat. {\bf 4}, 1281 (1992).
\bibitem{aliy} V.A.~Aliyev, G.D. ~Guseinov, F.I.~Mamedov, and L.M.
               ~Chapanova, Solid State Commun. {\bf 59}, 745 (1986).
\bibitem{choi} I.H.~Choi and P.Y.~Yu, Phys. Rev. B {\bf 63},
               235210 (2001).
\bibitem{vis} S.~Visbeck, T.~Hannappel, M.~Zorn, J.-T.~Zettler, and
              F.~Willig, Phys. Rev. B {\bf 63}, 245303 (2001).
\end{thebibliography}
\end{document}